\documentclass[preprint2]{aastex}

\slugcomment{\it To be submitted to the Astrophysical Journal}
\shortauthors{Yan et al.}
\shorttitle{Spitzer Mid-Infrared Spectroscopy of $z \sim 2$ ULIRGs}

\def\deg{\ifmmode {^{\circ}}\else {$^\circ$}\fi}
\def\kms{\ifmmode {\rm\,km\,s^{-1}}\else
    ${\rm\,km\,s^{-1}}$\fi}
\def\ergcm2s{\ifmmode {\rm\,ergs\,cm^{-2}\,s^{-1}}\else
    ${\rm\,ergs\,cm^{-2}\,s^{-1}}$\fi}
\def\ergAcm2s{\ifmmode {\rm\,ergs\,cm^{-2}\,s^{-1}\,\AA^{-1}}\else
    ${\rm\,ergs\,cm^{-2}\,s^{-1}\,\AA^{-1}}$\fi}
\def\ergs{\ifmmode {\rm\,ergs\,s^{-1}}\else
    ${\rm\,ergs\,s^{-1}}$\fi}
\def\kmsMpc{\ifmmode {\rm\,km\,s^{-1}\,Mpc^{-1}}\else
    ${\rm\,km\,s^{-1}\,Mpc^{-1}}$\fi}

\def\spose#1{\hbox to 0pt{#1\hss}}
\def\simlt{\mathrel{\spose{\lower 3pt\hbox{$\mathchar"218$}}
     \raise 2.0pt\hbox{$\mathchar"13C$}}}
\def\simgt{\mathrel{\spose{\lower 3pt\hbox{$\mathchar"218$}}
     \raise 2.0pt\hbox{$\mathchar"13E$}}}
     
\def\gs{\mathrel{\raise0.35ex\hbox{$\scriptstyle >$}\kern-0.6em
\lower0.40ex\hbox{{$\scriptstyle \sim$}}}}
\def\ls{\mathrel{\raise0.35ex\hbox{$\scriptstyle <$}\kern-0.6em
\lower0.40ex\hbox{{$\scriptstyle \sim$}}}}

\newcommand{\um}{\,$\mu$m}

\newcommand{\iso}{{\sl ISO }}
\newcommand{\spitz}{{\sl Spitzer }}
\newcommand{\zr}[2]{$z$\,$\sim$\,#1\,--\,#2}
\newcommand{\lir}[1]{10$^{#1}$\,$\rm{L}_{\odot}$}     

\def\plotfiddle#1#2#3#4#5#6#7{\centering \leavevmode
\vbox to#2{\rule{0pt}{#2}}
\includegraphics{#1}}

\begin{document}

\title{{\bf \it Spitzer} Mid-Infrared Spectroscopy of Infrared Luminous Galaxies at $z\sim2$ I: the Spectra}

\author{Lin Yan, Anna Sajina, Dario Fadda \\
Phil Choi, Lee Armus, George Helou, Harry Teplitz, David Frayer, Jason Surace}
\affil{\spitz Science Center, California Institute
of Technology, MS 220-6, Pasadena, CA 91125}

\email{lyan@ipac.caltech.edu}

\begin{abstract}
We present the mid-infrared (MIR) spectra obtained with the \spitz InfraRed Spectrograph (IRS) for a sample of 52 sources, selected as infrared luminous, $z\simgt1$ candidates in the Extragalactic First Look Survey (XFLS). The sample selection criteria are $f_{24\mu m} \simgt 0.9$mJy, $\nu f_\nu(24\mu m)/\nu f_\nu(8\mu m) \simgt 3.16$ and $\nu f_\nu(24\mu m)/\nu f_\nu(0.7\mu m) \simgt 10$. Of the 52 spectra, 47 ($90\%$) produced measurable redshifts based solely on the mid-IR spectral features, with the majority ($35/47=74\%$) at $1.5 \simlt z \simlt 3.2$. Keck spectroscopy of a sub-sample (17/47) agrees with the mid-IR redshift measurements. The observed spectra fall crudely into three categories --- {\bf (1)} 33\%\ (17/52) have strong PAH emission, and are probably powered by star formation with total IR luminosity roughly a factor of 5 higher than the local starburst ULIRGs. {\bf (2)} 33\%\ (17/52) have {\it only} deep silicate absorption at 9.8\um, indicative of deeply embedded dusty systems. The energetic nature of the heating sources in these systems can not be determined by these data alone. {\bf (3)} The remainder 34\%\ are mid-IR continuum dominated systems with either weak PAH emission and/or silicate absorption. This third of the sample are probably AGNs. We derived monochromatic, rest-frame 5.8\um, continuum luminosities ($\nu L_\nu$), ranging from $10^{10.3} - 10^{12.6} L_\odot$.  Our spectra have MIR slope $\alpha_{5-15\mu m} \simgt 2.1$, much redder than the median value of $1.3$ for the optically selected, Palomar-Green (PG) quasars. From the silicate absorption feature, we estimate that roughly two-thirds of the sample have optical depth $\tau_{9.8\mu m} > 1$.  Their $L_{1600\AA}$ and $L_{\rm IR}$ suggest that our sample is among the most luminous and most dust enshrouded systems of its epoch. Our study has revealed a significant population of dust enshrouded galaxies at $z\sim2$, whose enormous energy output, comparable to that of quasars, is generated by AGN as well as starburst. This IR luminous population has very little overlap with sub-mm and UV-selected populations.  
 
\end{abstract}


\section{Introduction}

Ultra-luminous Infrared Galaxies (ULIRGs, $L_{8-1000\mu m}$\,$>$\,\lir{12}), while relatively rare in the local universe \citep{soifer87,sanders96}, are far more common at high redshifts, as suggested by observations from \iso, (sub)-millimeter, and more recently \spitz \citep{Elbaz99, scott03, lefloch05, daddi05}. The tremendous energy output from ULIRGs makes them significant contributors to global luminosity density \citep{Guiderdoni98, dole06}. Detailed studies have found evidence for the evolutionary connections between ULIRGs and the formation of quasars and elliptical galaxies \citep{kalliopi06,veilleux06,alex05}.

The unprecedented sensitivity of the InfraRed Spectrograph \citep[IRS;][]{houck04} on \spitz has made it possible for the first time to observe galaxies at $z$\,$>$\,0.5$-$3, a virtually unexplored territory for mid-IR spectroscopy. The rest-frame $3 - 25$\um\ wavelength region is home to a rich suite of spectral features, essential for understanding the physical properties of ULIRGs. Low resolution, mid-IR (3\,--\,40\um) spectra of Luminous Infrared Galaxies (LIRGs) and ULIRGs consist of broad Aromatic emission (3.3, 6.2, 7.7, 8.6, 11.2, 12.7\um) and silicate absorption at 9.7\um\, plus various strong ionic or molecular lines, such as [ArII], [NeII], [NeIII], H$_2$, CO etc., superposed on a red continuum \citep{houck04,armus04,brandl04,spoon04}. The Aromatic features are usually ascribed to Polycyclic Aromatic Hydrocarbons (PAH), with $<$\,200 atoms and fluctuating effective temperatures responding to stochastic heating \citep{puget89}. These PAH features are prominent in star-forming systems, but are reduced and modified in high-intensity starbursts, eventually disappearing in AGN systems. They provide both an indication of the energy source heating the dust and redshift estimates for sources that are completely obscured at shorter wavelengths \citep{genzel00,draine03,laurent00,tran01,dim99,voit92}. The mid-IR continuum is associated with very small dust grains, which are more resilient to photo-destruction than the PAHs (Draine \&\ Li 2001). The mid to far-IR continuum ratio constrains the amounts of hot and cold dust. Finally, the depth of silicate absorption is a quantitative indicator of the mid-IR dust opacity along the line-of-sight. 

In this paper, we report the observational results from the \spitz GO-1 program (PID: 3748), which provides an initial characterization of mid-IR (7\,--\,38\um), low resolution spectroscopic properties of a sample of high-redshift, infrared luminous galaxies selected in the Extragalactic First Look Survey (XFLS). This program aims to address the basic spectral properties of high-z, IR luminous galaxies. This paper is the first in a series of papers from this survey. Its main scope is to present the target selection, observations (\S 2), spectral reduction method (\S 3), observed spectra and redshift distribution (\S 4). We discuss the implication of our findings in \S 5. The detailed analyses of the spectral properties --- PAH strength, mid-IR opacities and continuum slopes --- and the cosmological implication for the high redshift infrared luminous populations are presented in two subsequent papers (Sajina et al. 2006a, PAPER II hereafter; Sajina et al. 2006b, in preparation). 

Throughout the paper, we adopt the standard $\Omega_M$\,=\,0.27, $\Omega_\Lambda$\,=\,0.73\ and $H_0$\,=\,71\kmsMpc\ cosmology.

\section{Target Selection and Observations}

\subsection{The Target Sample}

We selected a total of 52 targets in the \spitz XFLS\footnote{For details of the XFLS observation plan and the data release, see http://ssc.spitzer.caltech.edu/fls.} over an area of 3.7deg$^{2}$, according to the following criteria: (1) $S_{24\mu m} \simgt 0.9$mJy; (2) $R(24,8) \equiv \log_{10}(\nu f_{\nu}(24\mu m)/\nu f_{\nu}(8\mu m) \simgt 0.5$; (3) $R(24,0.7) \equiv \log_{10}(\nu f_{\nu}(24\mu m)/\nu f_{\nu}(0.7\mu m) \simgt 1.0$. We used the \spitz and $R$-band catalogs published in Lacy et al. (2005), Fadda et al. (2006), and Fadda et al. (2004). The 24\um\ flux density of 0.9mJy cut off was chosen so that moderate signal-to-noise ratio ($S/N$) spectra could be obtained in a reasonable amount of telescope time. Using the infrared spectral template \citep{chary01}, the 0.9\,mJy at 24\um\ limit crudely translates to $L_{IR} = L_{1-1000\mu m} \sim 10^{12}L_\odot$ and $10^{13}L_\odot$ at $z=1$ and $2$ respectively. As discussed in detail in \S 4.4, the high luminosity of our sources determines the spectral properties observed among our sample.
 
\begin{figure}[!ht]
\begin{center}
\plotfiddle{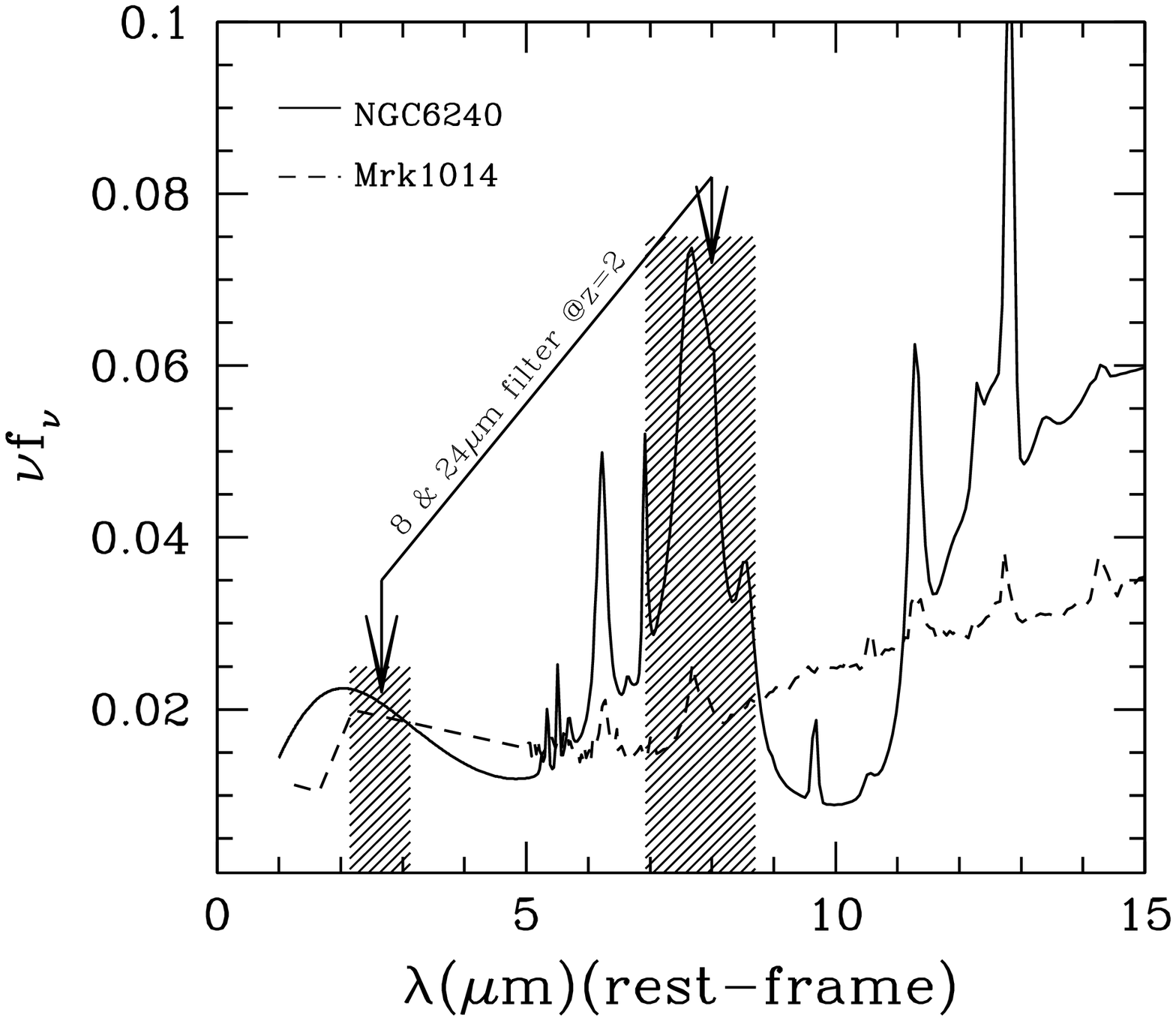}{3.0in}{0}{50}{50}{-160}{-115}
\end{center}
\caption{This plot illustrates how 24-to-8\um\ color can be used to select sources with PAH emission. It shows the mid-IR SEDs in the rest-frame wavelength in micron versus $\nu f_\nu$ for two ULIRGs which have been observed by the \spitz IRS \citep{armus04, armus06a}. Here we extended the SEDs down to 2~micron using the 2MASS and HST/NICMOS (NGC6240) broad photometry \citep{nick00}. The Y-axis $\nu f_\nu$ is arbitarily scaled. The shaded regions show the widths of the 8\um\ and 24\um\ filters at the rest-frame for $z=2$. The shaded areas mark the rest-frame wavelength of the 8 and 24\um\ filters at $z=2$. The monochromatic luminosity ratio between 24 and 8 micron ({\it i.e.}$R(24,8) \equiv \nu f_\nu(24\mu m)/\nu f_\nu(8\mu m)$) is redder for starburst dominated ULIRGs like NGC6240 than AGN dominated ULIRGs like Mrk1014 at $z \sim 2$.}
\label{sed}
\end{figure}

The 24-to-8\um\ and 24-to-R color cuts are to select galaxies with strong PAH or steep mid-IR continuum at $z\simgt 1$. This color selection technique has been discussed in detail by Yan et al.(2004), and more recently, by Brand et al. (2006). The $R(24,0.7) \simgt 1$ color cut corresponds to rest-frame $\nu f_\nu(8\mu m)/\nu f_\nu(2333\AA) \simgt 10$ at $z\sim2$, selecting infrared bright and optically faint sources. At $f_\nu (24\mu m) = 1$mJy, $R(24,0.7) \simgt 1$ implies $R$ magnitude fainter than $22.6$ magnitude (Vega). The 24-to-8\um\ color criterion picks out sources with strong PAH and/or steeply rising mid-IR continuum. $R(24,8) \simgt 0.5$ corresponds to $\nu f_\nu(8\mu m)/\nu f_\nu(2.7\mu m) \simgt 3.16$ at $z\sim2$, and the mid-IR slope $\alpha \simgt 2.1$ if $f_\nu \propto \nu^{-\alpha}$. Figure~\ref{sed} illustrates how this color selection works with two simple example spectra. At $z\approx 1 $ and $z \approx 2 $, the $R(24,8)$ ratio of a starburst ULIRG with strong PAH (NGC6240) \citep{armus06a} is redder than that of an AGN (Mrk1014) \citep{armus04}. This is due to the fact that strong, broad PAH emission near 8 and 12\um\ moves into the 24\um\ filter at $z \sim 2-1$. Obviously, this selection also picks up AGNs without much PAH emission but with steep red continua. 

\begin{figure}[!ht]
\begin{center}
\epsscale{1.0}
\plotone{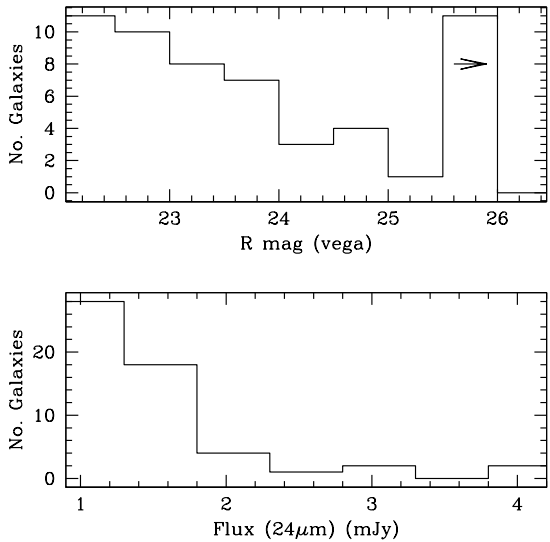}
\end{center}
\caption{These two panels show the distributions of 24\um\ flux density and
$R$-magnitude for our sample. The last bin with the arrow in the $R$ distribution plot shows the sources not detected in 
the $R$ images, and fainter than the $R$ limiting magnitude of 25.5. }
\label{fluxdist}
\end{figure}

Photometry for all targets in the three filters, 24\um, 8\um\ and $R$, are tabulated in Table~\ref{target_tab}. The final photometry is slightly different from what was initially available for the GO-1 target selection during the early period of the XFLS data reduction (see Figure~\ref{colortype} in Section 4). The change in MIPS 24\um\ fluxes is very minor, 5\%\ or less. The bigger change is in IRAC 8\um\ fluxes, partly because of the new and more accurate aperture corrections, and partly because most of our targets are very faint or not detected. This explains why $R(24,8)$ values for some of the targets are slightly smaller than $0.5$, a cut initially set by the early IRAC photometry. The full analyses of the selection function using Monte Carlo simulations will be performed in a separate paper (Sajina et al. 2006b). The brightness distributions at 24\um\ and $R$-band for our target sample are shown in Figure~\ref{fluxdist}.  The majority of our targets are 1mJy sources and the sample has a fairly narrow dynamic range in 24\um\ fluxes. The sources have $R > 22$, and many of them are fainter than the $R$-band limit of 25.5 magnitude (shown as the last bin in the $R$ distribution plot). The faint optical magnitudes motivated in part this survey to measure redshifts using solely the mid-IR spectra.

\subsection{The IRS Observations}

All observations were taken using IRS Staring mode, with nearby bright stars as peak-up targets to accurately center the slit on the science targets. For the 52 targets, we obtain low resolution (${\lambda \over \Delta \lambda} = 64 - 128$) spectra in the long-low module of the IRS (LL, $14 - 40\mu$m). For nine sources with IRAC 8\um\ fluxes greater than $150\mu$Jy, we also obtained Short-low, 1st order spectra (SL, $7.5 - 14\mu$m). For IRS low resolution staring mode, one cycle consists of two exposures at two different positions on the slit (two nods). 
For our sample, one cycle of 242~seconds at each nod position is taken for the SL 1st order observations, and three to nine cycles of 122~seconds were used for the LL, depending on the brightness of the source. Table~\ref{target_tab} shows the MIPS 24\um\ ID, source ID named after the IAU convention, the broad band 24\um, 8\um\ and R-band fluxes for the 52 sources. The source ID following the IAU convention provides the information on the source equatorial positions. We also listed the integration time of each individual exposure, the number of repeats at each nod position for SL 1st order and LL 1st and 2nd order, which can be used
to compute the total integration time for each order.  For completeness, we include the initial eight sources whose spectra have been published in \citet{yan05}. Our targets are all unresolved in all IRS wavelength. Since the IRS slit width is $3.7^{''}$ and $10.7^{''}$ for the SL and LL respectively, and the full-width-half-maximum (FWHM) at 8\um\ and 24\um\ is $\sim 2^{''}$ and $\sim 6^{''}$, light loss correction due to the finite IRS slit width is not needed for our observations. 

\section{Data Reduction and Analyses}

\subsection{Two-Dimensional (2D) Spectral Imaging Processing}

Our reduction starts with the 2D Basic Calibrated Data (BCD) produced by the IRS pipeline (S13 version) at the \spitz Science Center (SSC). The processing steps taken by the IRS pipeline include ramp fitting, dark sky subtraction, droop correction, linearity correction, flat fielding, and wavelength and flux calibration. We perform additional processing, including subtraction of the ``residual'' background in the BCD images, masking any significantly deviant pixels, and stacking images taken at the same nod position. We build the background image by using only the BCD images of the same target, including both nod positions. At the LL, we have roughly (12-32) BCD images for computing the background. Our method is iterative. In the first pass, a median combined background image is obtained after masking out the target spectra. After subtracting this background image from each BCD image, we go through the second iteration of identifying and masking out serendipitous spectra in the images, then produce the final background estimate. To minimize the effects of the deviant pixels, we used a biweight estimator \footnote{reference: "Data Analysis and Regression: A Second Course in Statistics", Mosteller and Tukey, Addison-Wesley, 1977, pp. 203-209.}  to compute the background and its dispersion. The noise in each pixel of the final images is computed by adding quadratically the noise evaluated from the background to the Poisson noise from the target spectra. The error spectrum is measured from the corresponding noise image. 

Before spectral extraction, we identify and correct the ``rogue pixels'' in the 2D images. The ``rogue pixels'' are pixels whose dark current is abnormally high and varies with time and different sky background radiation. These pixels are highly deviant from the local background. We identify the ``rogue pixels'' by computing the dispersion of the noise around every pixel and flagging pixels which are above 5$\sigma$ of the mean value. This threshold is relatively conservative. For any deviant pixel which {\it does not} fall on the spectra (within one FWHM profile), the code automatically sets its value to zero (which is the value after background subtraction). For the deviant pixels which {\it do} fall on the target spectra (within one FWHM), we visually examine each pixel, and replace its value with the linear interpolation ($\pm1-2$ pixels) along the wavelength direction. The interpolation is appropriate for our data because most of the detected spectral features ({\it i.e.} PAH) are much broader than 1-2 pixels, and will not be smeared significantly. This type of correction is usually needed for only one or two pixels per spectrum. 

\onecolumn

\begin{figure}
\epsscale{1.1}
\plotone{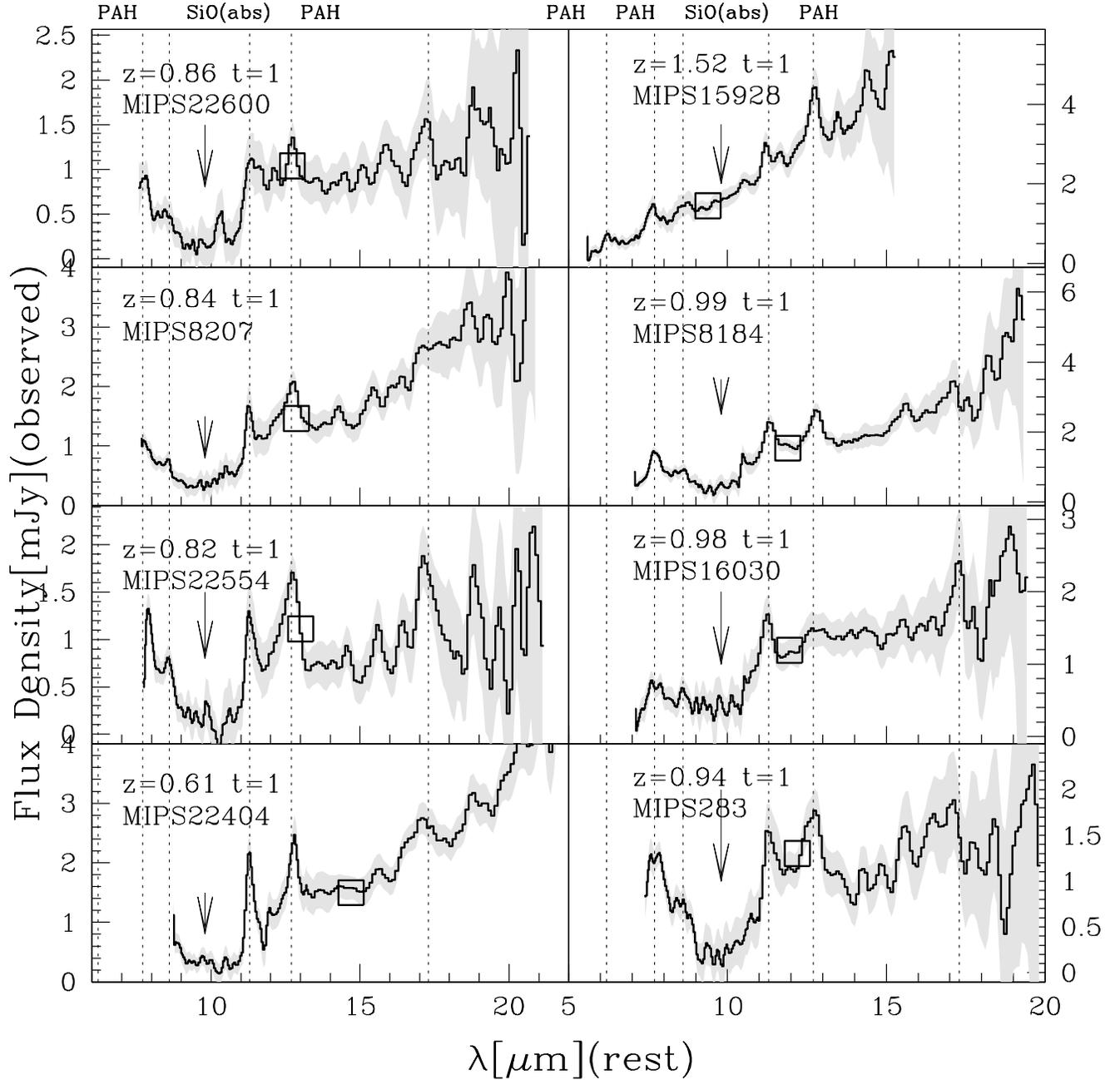}
\caption{The mid-IR spectra for 47 sources with redshift measurements. The spectra are presented in observed flux density $f_\nu$ in mJy versus the rest-frame wavelength in micron. The spectra were smoothed by a $3$~pixel boxcar, in order to enhance the broad features such as PAH emission and silicate absorption. In each panel, the redshift, source ID and the spectral type (see text for the description) are labelled at the top-left corner. The large, open squares indicate the broad band, 24\um\ photometry, and the gray shaded region marks the $\pm 1\sigma$ error for the spectrum. The error spectrum is not smoothed. The error computation is discussed in \S 3.2.  The dashed lines mark the PAH emission features, and the arrows mark the silicate absorption. \label{spec}}
\end{figure}

\addtocounter{figure}{-1}
\begin{figure}[!t]
\epsscale{1.1}
\plotone{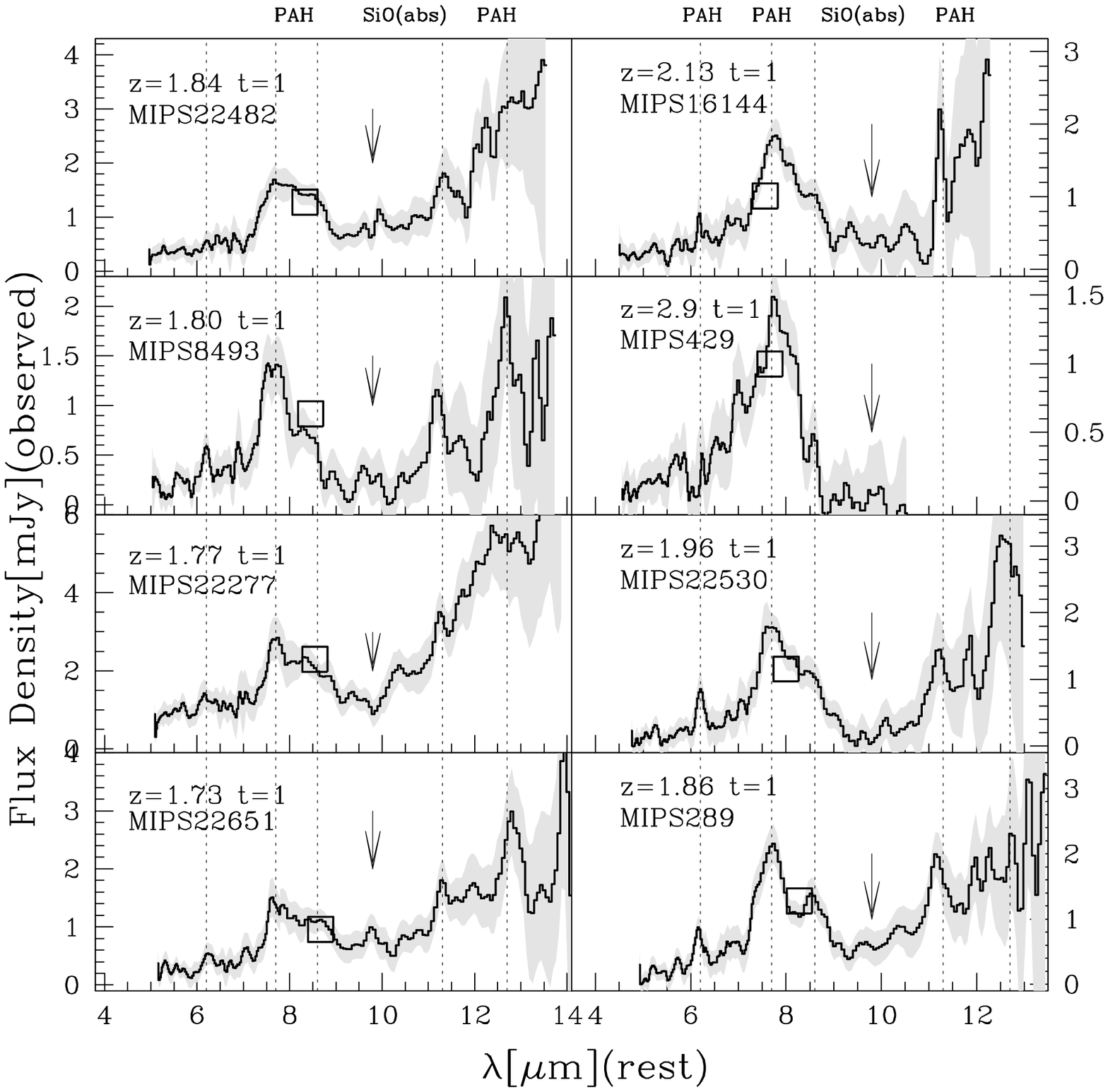}
\caption{Continue.}
\end{figure}

\addtocounter{figure}{-1}
\begin{figure}[!t]
\epsscale{1.1}
\plotone{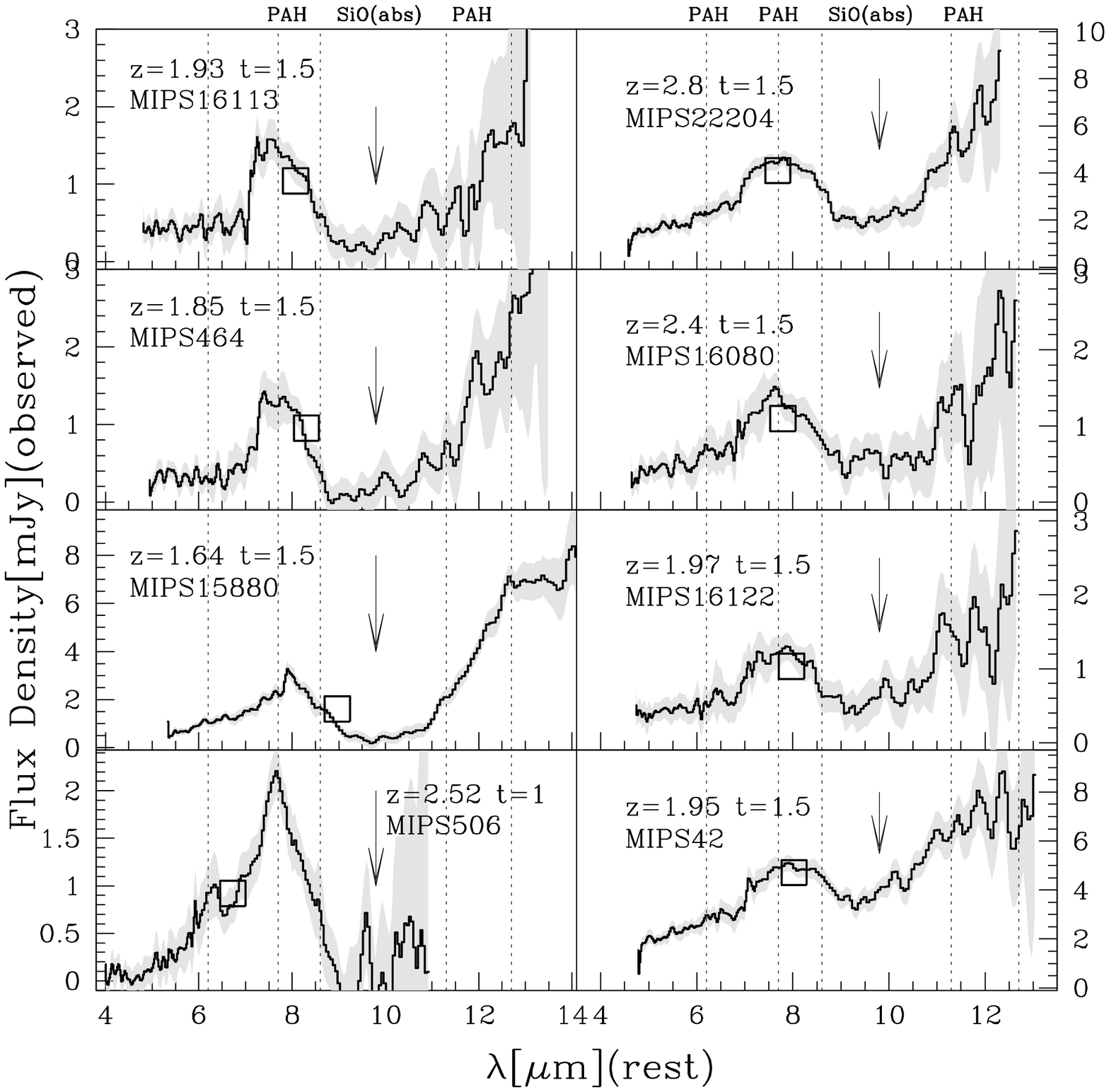}
\caption{Continue.}
\end{figure}

\addtocounter{figure}{-1}
\begin{figure}[!t]
\epsscale{1.1}
\plotone{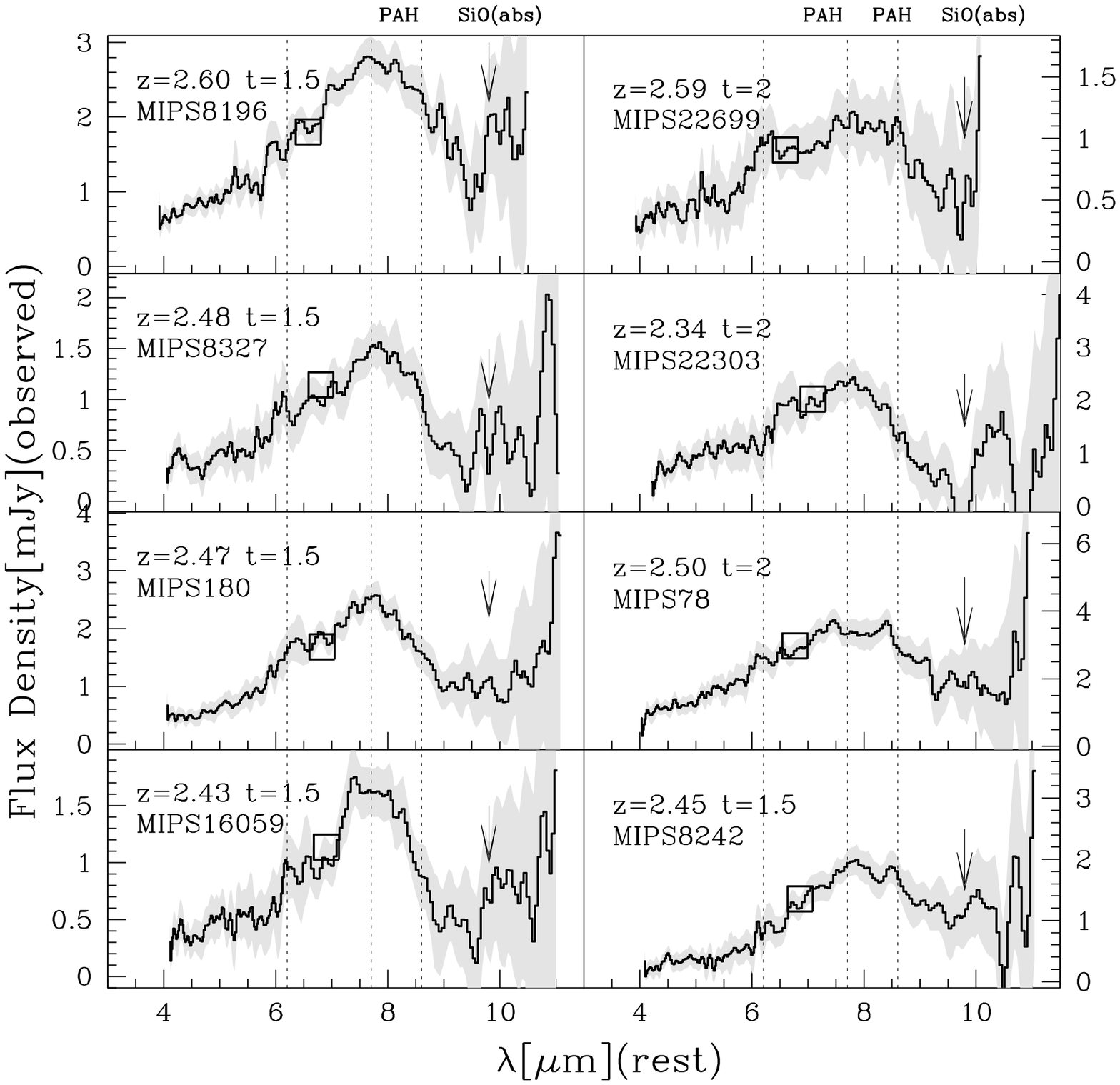}
\caption{Continue.}
\end{figure}

\addtocounter{figure}{-1}
\begin{figure}[!t]
\epsscale{1.1}
\plotone{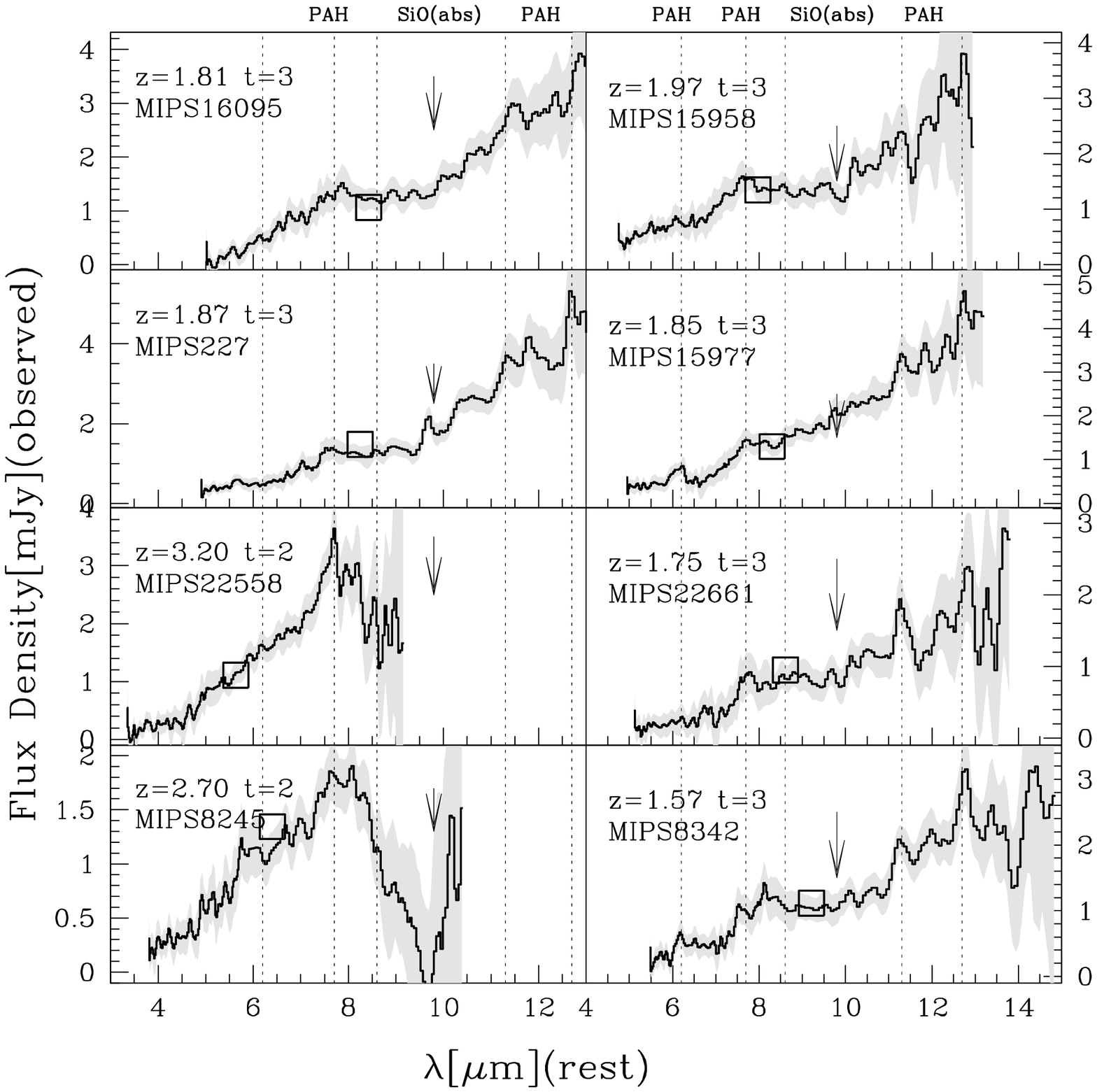}
\caption{Continue.}
\end{figure}

\addtocounter{figure}{-1}
\begin{figure}[!t]
\epsscale{1.1}
\plotone{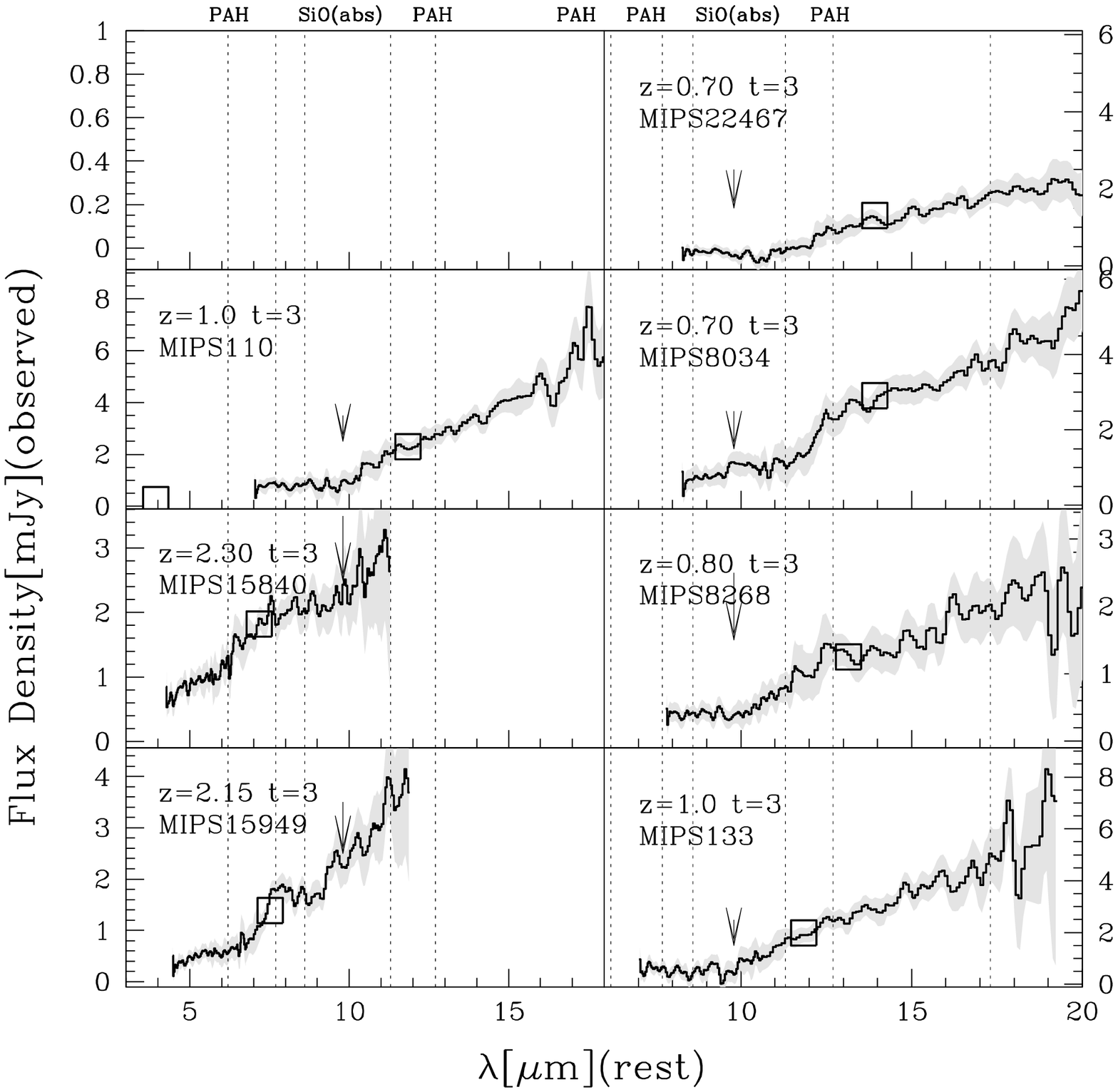}
\caption{Continue.}
\end{figure}

\begin{figure}[!t]
\epsscale{1.1}
\plotone{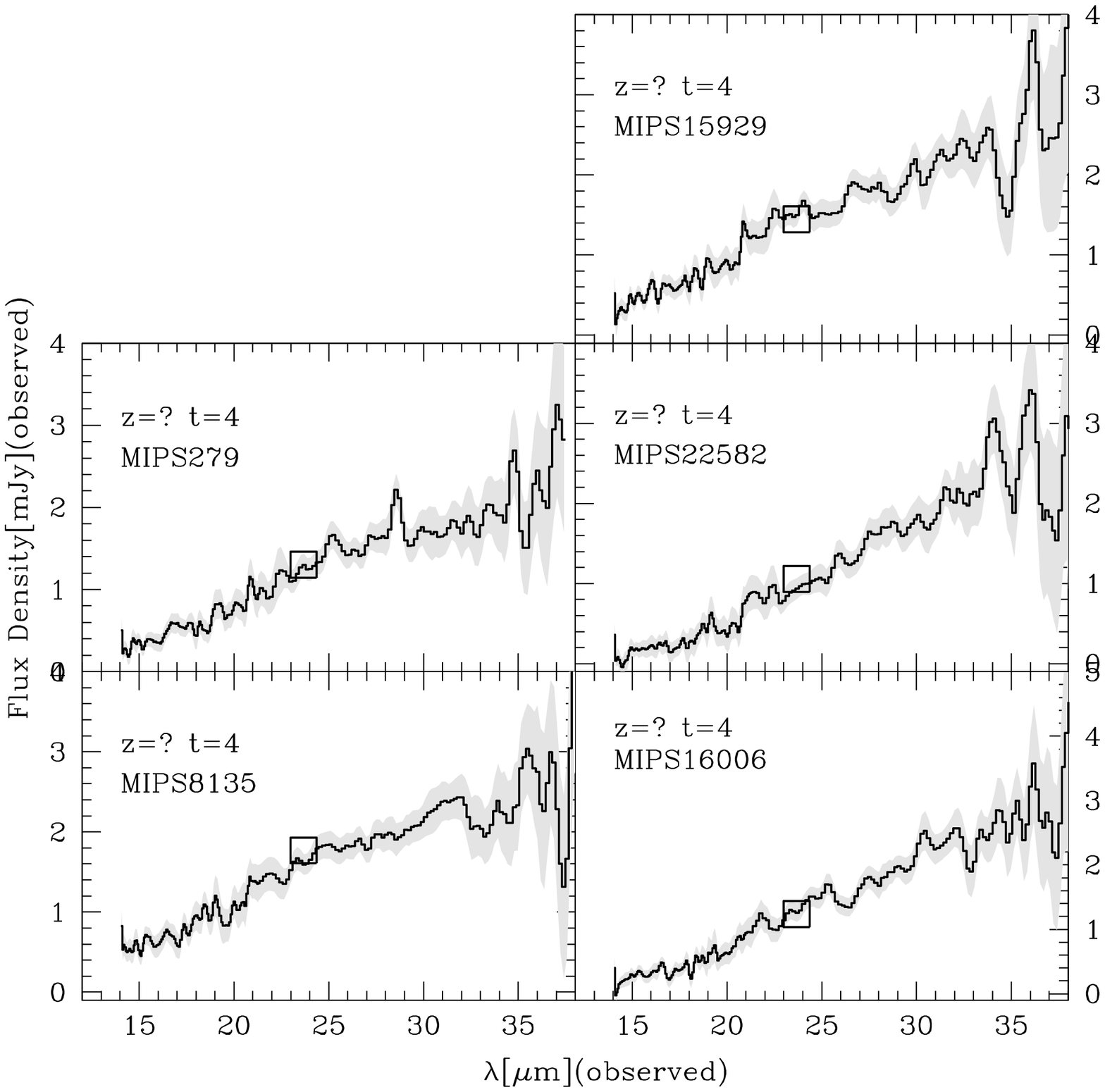}
\caption{The five observed spectra that did not yield any redshift measurements. The spectra are plotted in the observed flux density versus observed wavelength. As in Fig~\ref{spec}, the open squares indicate 24\um\ broad band photometry. The five spectra have power-law continua without identifiable spectral features. The spectral type is set to $t=4$. \label{nozspec}}
\end{figure}

\twocolumn

\subsection{Spectral Extraction}

Using the steps described in the section above, a final coadded 2D spectral image is produced for each nod position for each IRS slit. One-dimensional (1D) spectra were extracted, then averaged to produce the final 1D spectrum for each module. The spectral extraction is done using the SSC software SPICE v1.3beta version\footnote{see http://ssc.spitzer.caltech.edu/postbcd/spice.html for the description of the software and the public release status.}. In SPICE, the spectral extraction aperture can be adjusted to maximize the $S/N$ ratio. For our data, we used an extraction aperture of 6~pixels at 27\um\, 1~pixel narrower than the default PSF size in SPICE, and the shape of the extraction aperture simply follows the PSF size as a function of wavelength. This method minimizes the uncertainties due to the background, and maximizes the $S/N$ ratios for faint spectra. We used the simple method of summing up the light within the extraction aperture. The flux scales of the final, averaged SL and LL spectra are consistent with the IRAC 8\um\ and MIPS 24\um\ broad band fluxes. No rescaling is applied to any of the spectra. The final spectrum for each source is produced by averaging the spectra from the two nod positions and combining the spectra from the 1st, 2nd and 3rd order together using linear interpolation. Different orders have different pixel-to-wavelength scales, therefore, we resample the overlapping spectra onto the smallest scale, and produce the final combined spectra by averaging. We found that the flux calibration between different orders is consistent.

\section{Results}

\subsection{Redshift Measurements from Mid-IR Spectra} 

The observed 52 spectra from the sample are presented in Figure~\ref{spec} and Figure~\ref{nozspec}.
Of the 52 sources, 47 have redshifts measurable from their mid-IR spectral features, such as PAH emission and silicate absorption, yielding redshift efficiency of 90\%. The remaining five sources have power-law spectra without any identifiable features, thus have no redshift measurements (Figure~\ref{nozspec}).  For the majority of our spectra, the redshifts are determined by the multiple spectral features, and are thus secure.  As in any spectroscopic redshift studies, for a few sources,  the spectral identifications could have some ambiguities. We therefore identify our redshift measurements by a quality flag $Q_z$, with $Q_z=a$ for secure identification, and $Q_z=b$ for redshifts spectral features whose identifications are ambiguous. We have only 6 $z$ with $Q_z=b$. If excluding these 6 redshifts, our redshift efficiency is 80\%. Flag $Q_z$ and redshift errors are two different parameters which quantify the security of the spectral identification and redshift accuracy respectively.

The redshift errors are estimated as follows. We compute individual $(1+z)_{i} = \lambda_{obs}(i)/\lambda_{rest}(i)$, using spectral feature $i$, based on the measured feature centroid $\lambda_{obs}(i)$ and the corresponding rest-frame wavelength $\lambda_{rest}(i)$. The final redshift $z$ is the average value of $z_{i}$, and the $1\sigma$ deviation is computed as $1\sigma = {{\sqrt{\sum_i (z_i - <z>)^{2}}} \over \sqrt{n-1}} $. Another source of redshift errors is $\Delta z_i$, from the centroid uncertainty $\Delta \lambda_{i}$ for each feature. The total $z$ error should be the quadratic sum of the $1\sigma$ standard deviation and $\Delta z_i$. For spectra whose redshifts based on a single spectral feature, the uncertainty is only estimated from the uncertainty of the feature centroid. For example, for spectra with only silicate absorption, the redshift errors can be as large as 0.1\,--\,0.2 because of the intrinsic broad width of this feature. For spectra with both PAH emission and silicate absorption features, we  use {\it only} PAH features for measuring redshifts since they are generally narrower than silicate absorption, thus give more accurate results and smaller redshift errors.

One of the critical parameters in determining the redshifts are the rest-frame wavelength $\lambda_{rest}(i)$ for various PAH features. These broad, complex emission bands roughly have similar rest-frame centers, although small variations have been found in studies of local PDRs (Vermeij et al. 2002; Peeters et al. 2004). For our purpose, we measure $\lambda_{rest}(i)$ from the mid-IR spectrum of a local starburst galaxy NGC7714, observed by IRS in the same configuration and resolution as used for our sample \citep{brandl04}. Taking into account the distance, we computed the rest-frame wavelengths of PAH emission features based on the measured centroids. We used the following values for the reference 
wavelength $\lambda_{rest}(i)$, $6.2180\pm0.0061, 7.7118\pm0.0229, 8.5972\pm0.0077, 11.2665\pm0.00353, 12.8030\pm0.0077$\um\ for 6.2,7.7,8.6,11.2 and 12.7\um\ PAH features respectively.

Table~\ref{ztable} lists the redshift, the redshift error, the estimated rest-frame luminosity at 5.8\um, $\nu L_{\nu}(5.8\mu) = L_{5.8\mu m}$, the redshift quality flag ($Q_z$), the spectral group classification ($t$), and the spectral features used for redshift measurements. 

\begin{figure}[!ht]
\plotone{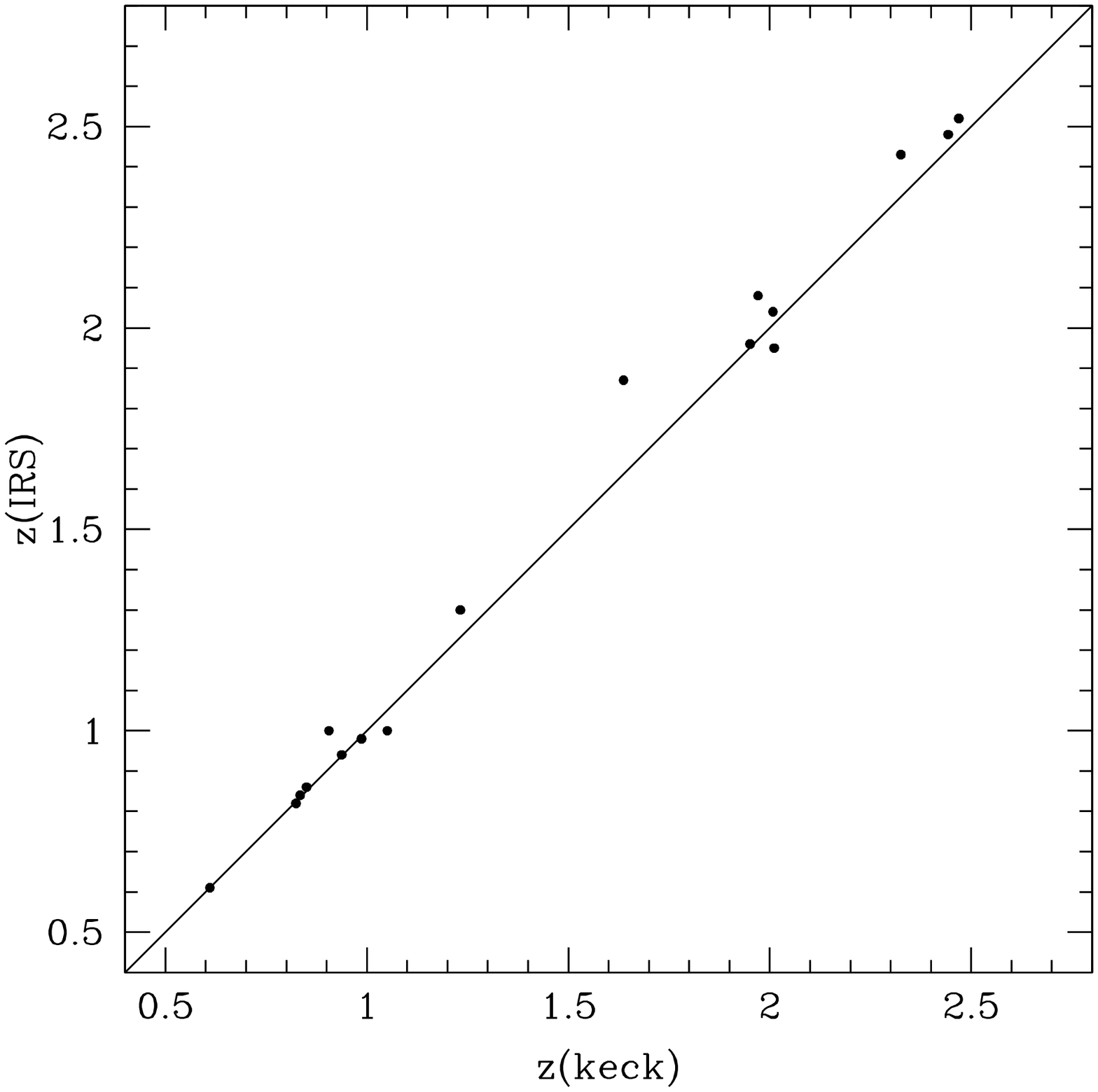}
\caption{We compare the redshifts measured from the Keck
optical and near-IR spectra with the IRS-based redshifts.
The straight line marks the $1:1$ ratio. \label{zkeck}}
\end{figure}

We obtained optical and near-IR spectra for a subset of the IRS sample using LRIS, DEIMOS and NIRSPEC on the Keck telescopes \citep{oke94,faber03,mclean00}. We were able to determine 17 redshifts from the Keck spectra. All of the Keck redshifts confirm the estimates based only on the mid-IR features, even in a few ambiguous cases where the IRS redshifts are based only on weak features. Figure~\ref{zkeck} illustrates the Keck $z$ versus the IRS-based $z$. For two sources with weak features, MIPS110 and MIPS133 with $t$\,=\,3 and $Q_z$\,=\,$b$, their IRS redshifts are primarily based on the mid-IR continuum slope change around 10\um. These results have been confirmed by their Keck spectra, suggesting that our visual identification of weak mid-IR features is reasonable. This 10\um\ break is probably not a new feature, but due to small amount of silicate absorption  at 9.8\um, and generally the slightly curved shape at the break supports this idea (see Figure~\ref{spec}). One of the five sources without IRS-based redshifts, MIPS279, was observed with the Keck LRIS, and the redshift is at 1.23.  Table~\ref{keckspec} shows the redshifts from the Keck and the IRS. We also listed the source type classified from the IRS, as well as from the optical spectra when it is possible.  At the near-IR, we have only enough wavelength coverage to detect $H_\alpha$, thus definitive source type classification is not possible. However, all $H_\alpha$ emission lines are narrower than 800~km/s. As shown, the redshift and the classification from the optical/near-IR data are consistent with what measured from the \spitz spectra.

\subsection{Mid-IR Spectral Type}

Examining Figure~\ref{spec} and Figure~\ref{nozspec}, it is apparent that the 52 spectra in our sample fall broadly into three categories: strong PAH emission sources, sources without PAH but with deep silicate absorption at 9.8\um\, and mid-IR continuum dominated sources with weak or no emission/absorption features. Making this crude classification according to broad spectral features allows us to have a quick assessment of the main properties of the sample. PAPER II of this series discusses in detail the full spectral fitting and analyses. We use numeric $t$\,=\,1 to designate systems with strong PAH emission. Quantitatively, the spectra classified by eye in this category roughly correspond to 7.7\um\ rest-frame equivalent width (EW) greater than 0.7\um, or equivalently, 6.2\um\ EW$_{rest} > 0.2$\um\ (computed in PAPER II).  Sources without any evidence for PAH emission, but strong 9.8\um\ silicate absorption feature are designated $t$\,=\,2. Intermediate sources between these two types, i.e. with strong Si absorption, as well as prominent 8\um\ bumps, which are possibly due to PAH emission (occasionally supported by 6.2\um\ features) are designated $t$\,=\,1.5.  The rationale for separating this sub-group ($t$\,=\,1.5) is that a good fraction of high-$z$, IR-luminous sources seem to have this type of spectra, weak PAH emission with strong silicate absorption \citep{houck05, weedman06}. The mid-IR continuum dominated galaxies are separated into sources with strong continuum plus weak PAH emission and/or silicate absorption features ($t$\,=\,3), and  sources with pure power-law continua without identifiable features ($t$\,=\,4, thus without redshifts). As shown in Figure~\ref{spec}, we have 17 sources with strong PAH emission ($t$\,=\,1), 17 sources with only deep silicate absorption (12 with $t$\,=\,1.5, 4 with $t$\,=\,2), 13 spectra with $t$\,=\,3 whose emission and absorption features are weak, and finally, 5 sources with power-law continua without redshifts.  Some of these five sources could be AGNs with very red mid-IR continua, and the rest could also be starbursts at $z>3$, where most strong PAH emission and silicate absorption features have redshifted out of the IRS longest wavelength window ($5 - 38$\um). 

\subsection{Redshift distribution versus spectral type}

\begin{figure}[!ht]
\begin{center}
\plotone{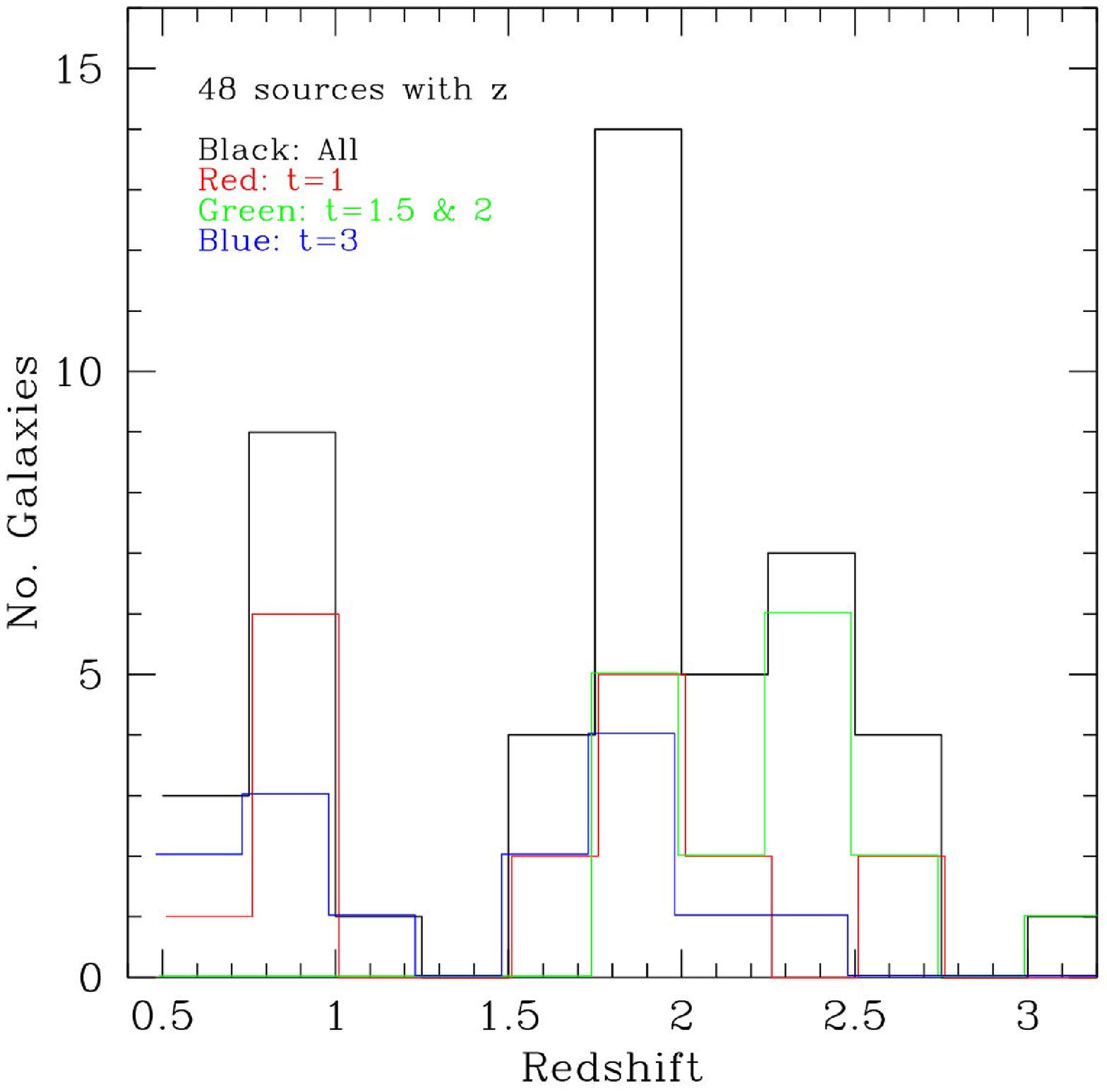}
\end{center}
\caption{Redshift distributions separated by the spectral types. The black histogram is for all of the redshifts (47), and the red is for 17 of the strong PAH sources ($t$\,=\,1), the green for 17 of the highly-obscured sources ($t$\,=\,1.5\,\&\,2), the blue for 13 of the continuum dominated objects with weak features (emission and/or absorption). } 
\label{ztype}
\end{figure}

Figure~\ref{ztype} presents the redshift distribution of all 47 redshifts, as well as the z-distribution for each source type. Of the 47 redshifts, 12 are $0.6 < z < 1.3$ and the remaining 35 are beyond $z = 1.5$.  This indicates that our selection criteria are indeed very effective in selecting $z\simgt$ IR luminous galaxies. The low-$z$ contamination comes from the fact that at $0.6 < z < 1$, IRAC 8\um\ and MIPS 24\um\ filters sample the rest-frame 4\um\ continuum and 12.7\um\ PAH respectively. The strong 12.7\um\ PAH emission in these low-$z$ sources boosts the 24/8\um\ color into our red color cut. The dip at $ 1.3 < z < 1.6$ in Figure~\ref{ztype} is due to the fact that at $z$\,=\,1.5, the 24\um\ filter samples the rest-frame 9.8\um, thus any samples with 24\um\ flux cutoff will inherently bias against selecting sources with 9.8\um\ silicate absorption around redshifts of $1.5$. Sources without strong silicate absorption should be selected. For example, we have two sources, MIPS15928 and MIPS8342, at $z=1.52$ and $1.57$ in our sample, which have weak or no 9.8\um\ absorption as shown in Figure~\ref{spec}.  As shown in Figure~\ref{ztype}, all of the 16 sources with {\it only} deep silicate absorption ($t$\,=\,1.5 and $t$\,=\,2) are at redshifts beyond $1.5$. Of the 17 sources with strong PAH ($t$\,=\,1), 7 are at $z$\,$\sim$\,0.61\,--\,1.0, and the remaining 10 are at $z$\,$\sim$\,1.52\,--\,2.59. For the 13 sources with $t$\,=\,3, 6 are at $z$\,$\sim$\,0.7\,--\,1.23, and 8 at $z$\,$\sim$\,1.57\,--\,2.4.

\subsection{Stacking the mid-IR Spectra}

\begin{figure*}
\begin{center}
\vspace*{8.5cm}
\leavevmode
\includegraphics{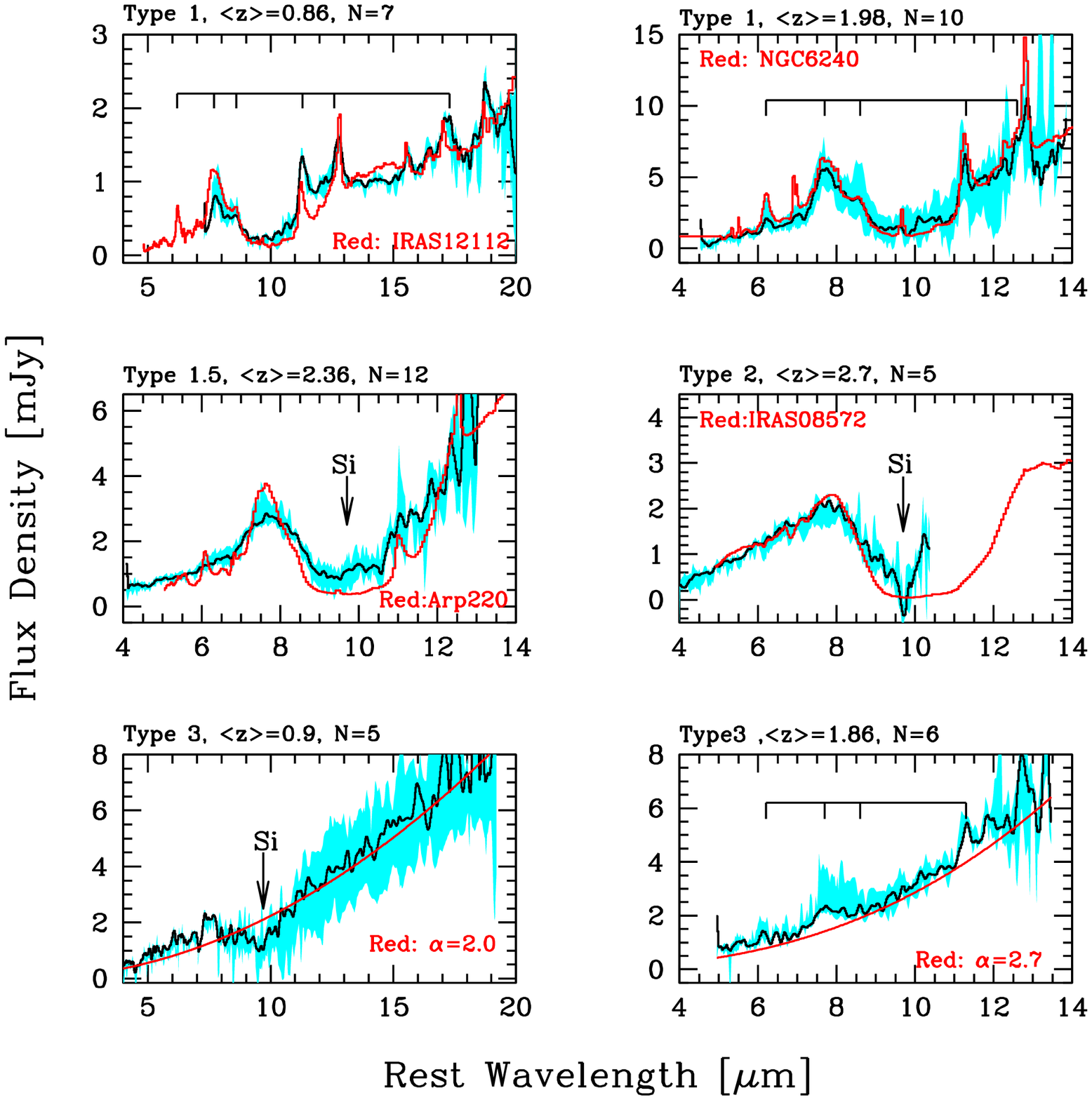}
\end{center}
\caption{The six panels show the median spectrum (in black line) per type for $z$\,$\simlt$\,1 and $z$\,$\simgt$\,1 sources. For type $t=1$ and $t=3$, we stacked spectra for $z$\,$\simlt$\,1 and $z$\, $\simgt$\,1 separately.  For class $t=3$ and $z>1$ bin, we excluded two spectra in the stacking (see the text for detail). Type $t=1.5$ and $t=2$ have only $z$\, $\simgt$\,1. The cyan shaded regions indicating the 20 and 80 percentfile. The source type, average $z$, and number of spectra included in the stacking are indicated at the top-left corner of each panel. The red spectra in the first four panels represent the scaled local ULIRG spectra to illustrate the similarity between spectra at high-z and $z=0$.  The local ULIRG spectra were chosen from Armus et al. (2006b) to have the closest match with the median spectra. In the last two panels, for $t=3$ class, the red lines indicate power-law continua with $\alpha$ ($f_\nu \propto \nu^{-\alpha}$) of  $2.2$ and $1.8$ for $z>1$ and $z<1$ respectively. The slope $\alpha$ is much steeper than that of the local AGN spectra like NGC1068.
\label{medspec}}
\end{figure*}

To increase the $S/N$ ratio and to examine the variation and uniformity of the spectral features, we applied stacking analyses to the spectra separated by each class and by redshift ($z<1$ and $z>1$). The spectra are normalized either at 4.8\,--\,5.2\um\ for sources at $z\simgt 1$, or at 15\um\ for $z\simlt 1$. We computed the median spectra as well as 20 and 80 percentile spectra to illustrate the spectral variation.  For the strong PAH class ($t$\,=\,1),  because of different rest-frame wavelength coverage, the spectra at $z$\,$\gs$\,1 and at $z$\,$\ls$\,1 are stacked separately. The same approach is taken for type $t=3$, except that we exclude two spectra in the $z>1$ bin due to their much steeper continua. The stacking results are shown in Figure~\ref{medspec}. The improved $S/N$ median spectra confirm the spectral features identified in individual spectrum. Particularly, for $t=3$, $z>1$ spectra, although the PAH features in individual spectrum are quite weak, their presence is indeed confirmed by the median stacked spectrum for this class. The median spectrum for $t=3, z<1$ objects shows the weak absorption trough at 9.8\um, although truncated, confirming the redshift determinations for this group of sources. 

Similar types of mid-IR spectra have been seen among local ULIRGs \citep{armus06b}. In Figure~\ref{medspec}, we overlay in red lines the rescaled, IRS spectra of local ULIRGs. We emphasize that these local spectra were plotted {\it not} for the purpose of spectral fitting, but to illustrate their similar shapes to our spectra at $z\sim1-2$. For example, our type $t=2$ spectra are similar to the local ULIRG IRAS F08572+3915, which is a highly embedded system \citep{spoon05}. Type $t=1.5$ spectrum has a broad peak around $7-8$\um. The closest example we can find among the local, available ULIRG spectrum is Arp\,220 \citep{armus06b}. It is possible that some of that broad $7-8$\um\ peak could be from PAH emission, and not just simple mid-IR continuum.  For $t=3$ spectra, we plot the pure power-law continua, indicated with red lines in the figure. The slopes ($\alpha$, $f_\nu \propto \nu^{-\alpha}$) are $2.2$ and $1.8$ for the median spectra at $z>1$ and $z<1$ respectively. These slopes are much steeper than the mid-IR spectra of classic AGN, like NGC1068, with $\alpha$\,$\sim$\,1.
 
Figure~\ref{medspec} illustrate three broad categories of spectra observed in our sample, spectra with clear PAH emission, ones with only deep silicate absorption and continuum dominated spectra with only weak PAH and/or silicate absorption. Taking the median spectra, we measured the optical depth at $9.8$\um, $\tau_{9.8\mu m}$\,=\,$\ln(S_{\rm{cont}}/S_{\rm{min}}$), by simply approximating the continuum with a power-law between 7\um\ and 13\um. This crude estimates show silicate absorption is significant for both strong PAH sources ($t=1$) as well as absorption only systems ($t=1.5, 2$). These spectra are optically thick at 9.8\um\ with $\tau_{9.8\mu m} \sim 1.1 - 1.3$, implying that close to 60\%\ of our sample are heavily obscured at mid-IR. We note that for systems with strong PAH emission, the measured optical depth at 9.8\um\ silicate absorption provides only a lower limit since the broad wings from 7.7, 8.6 and 11.3\um\ PAH emission tend to quickly fill in the absorption trough, resulting under-estimate of the true opacity. 

\subsection{Luminosities}
\subsubsection{5.8\um\ monochromatic continuum luminosity \label{contlums}}

\begin{figure}[!ht]
\begin{center}
\plotone{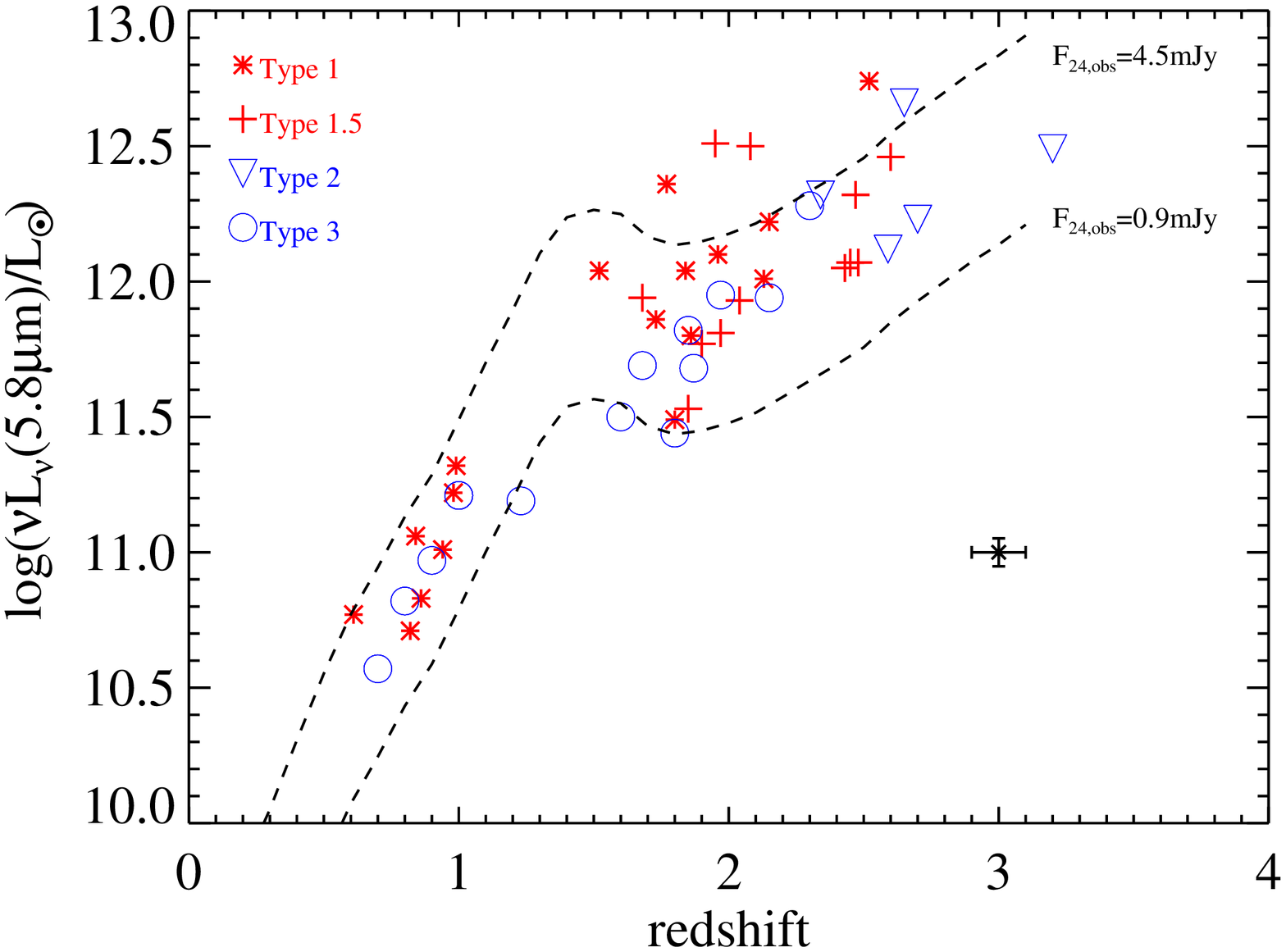}
\end{center}
\caption{Monochromatic, rest-frame 5.8\um\ continuum luminosity vs. redshift with different
symbols marking the different source types. The red and blue color crudely separates the sources with strong and weak PAH emission respectively. The dashed line shows the $L_{5.8}$ vs. $z$ relation if we take a SED like NGC6240 with the $F_{obs}(24\mu m)$ scaled to the lower and upper limits of our survey sample. The errorbar shows the mean errors for our sample. 
\label{lumtype} }
\end{figure}

To characterize the energetic output of our sources, we measure the rest-frame 5.8\um\ monochromatic, continuum luminosity, $L_{5.8} = \nu L_\nu (5.8\mu m)$ from our spectra. This is the rest-frame wavelength where we have data for most of our sources. Table~\ref{ztable} includes the derived $\log L_{5.8}$  luminosities for our sample, which range between $\sim$\,10.3 for the lowest-$z$ sources up to $\sim$\,12.6.  Figure~\ref{lumtype} shows the distribution of $L_{5.8}$ versus $z$ for our sample with sources with and without strong PAH emission in red and blue symbols respectively. We found that some of the strong PAH emitting galaxies at $z\sim2$ are $\sim$\,0.5\,dex more luminous than local starburst ULIRGs \citep{armus06b}.  The mid-IR luminosity trend with redshift is essentially our selection function. The minimum and maximum 24\um\ fluxes are translated to $L_{5.8}$ using NGC6240 spectrum, shown as the dashed lines in the figure.  The peak and the valley at $z\sim1.5$ and $z\sim2$ reflect the silicate absorption trough and the strong 7.7\um\ PAH emission passing through the observed 24\um\ filter at these redshifts.  Studies of local ULIRGs suggest that at $L_{\rm IR} = L_{8-1000\mu m} \sim$\,10$^{12.5}$L$_{\odot}$, ULIRGs tend to be AGN-dominated \citep{lutz98,veilleux99}. In Figure~\ref{lumtype}, we do not observe any significant trend in mid-IR luminosity versus source type. This does not imply that our high-z sample is different from the local ULIRGs.  To study the luminosity versus source type, we need to derive accurate $L_{\rm{IR}}$ using long wavelength data at 1.2mm \citep{lutz05b} and MIPS 70, 160\um.  We defer this analysis to a later paper.

However, in order to put our sources in context with other galaxy populations, here we use a crude estimate of $L_{\rm{IR}}$ based on $L_{5.8}$.  This is necessarily very crude, as 5.8\um\ does not directly sample the far-IR thermal emission dominating the bolometric infrared emission.  Brandl et al. (2006) showed that using the 15\um\ and 30\um\ fluxes together addresses both these issues, and provides a good conversion to $L_{\rm{IR}}$. For the $z$\,$\ls$\,1.6 sources (about 1/3 of the sample), using both the IRS spectra and the MIPS70\um\ points we can derive both these points (see PAPERII for details).  Using these and the Brandl et al. relation, we find and average $L_{\rm{IR}}/L_{5.8}$\,$\sim$\,$25^{+10}_{-8}$. This average ratio is consistent with the value derived from the IRS spectra of the local ULIRGs sample, except for the extremely cold and warm outliers Arp220 and IRAS08572+3915 (for this source, the ratio is the lowest, 5) \citep{armus06b}.  Given the 5.8\um\ luminosities in Table~2 and the above ratio implies that our sources have $L_{\rm{IR}} \sim 10^{12}-10^{13}L_\odot$ and are ULIRGs or type II QSOs. With the average of $<L_{5.8}>=9\times10^{11}L_\odot$, even with the lowest ratio of 5, the implied $L_{\rm {IR}}$ is still $4.5\times10^{12}L_\odot = 2\times10^{46}$~ergs/s, similar to the energy output from quasars.

\subsubsection{The PAH fraction of mid-IR luminosity}
In normal star-forming galaxies, PAH emission accounts for $\sim$\,10-15\% of the total infrared luminosities \citep{smith06}. For local ULIRGs, this fraction is small, on the order of a few percent (Armus et al. 2006b). Here we give a crude estimate by measuring the PAH fraction in the median spectra with strong PAH (see Fig.~\ref{medspec}). A more careful spectral fitting of individual spectrum is done in PAPERII.  We estimate the fraction of summed PAH luminosity over 5.8\um, monochromatic continuum luminosity, $\sum L_{\rm{PAH}} / \nu L_\nu (5.8\mu m)$. The sum of the PAH luminosities, $\sum L_{\rm{\rm{PAH}}}$, includes the 6.2, 7.7, 8.6 and 11.3\um\ features.  We use a simple straight-line continuum between 5\um\ and 14\um, which allows for a Si absorption feature at 10\um. The PAH flux is estimated by integrating the full features from the continuum-subtracted spectra.  Given the 20th and 80th percentile range in the median spectra shown in Fig.~\ref{medspec}, we obtain $\sum L_{\rm{PAH}} / \nu L_\nu (5.8\mu m)$\,$\sim$\,0.2\,--\,0.7. This implies that the PAH emission accounts for $\sim$\,1\,--\,3\% of the total infrared luminosities ($L_{\rm{IR}}$), which is comparable to that of local ULIRGs \citep{armus06b}, but lower than pure starburst galaxies from Brandl et al. (2006).

This is much smaller than what is measured among normal galaxies, and similar to that of local ULIRGs.   

\subsubsection{The rest-frame UV luminosities}
\begin{figure}[!h]
\begin{center}
\epsscale{1.}
\plotone{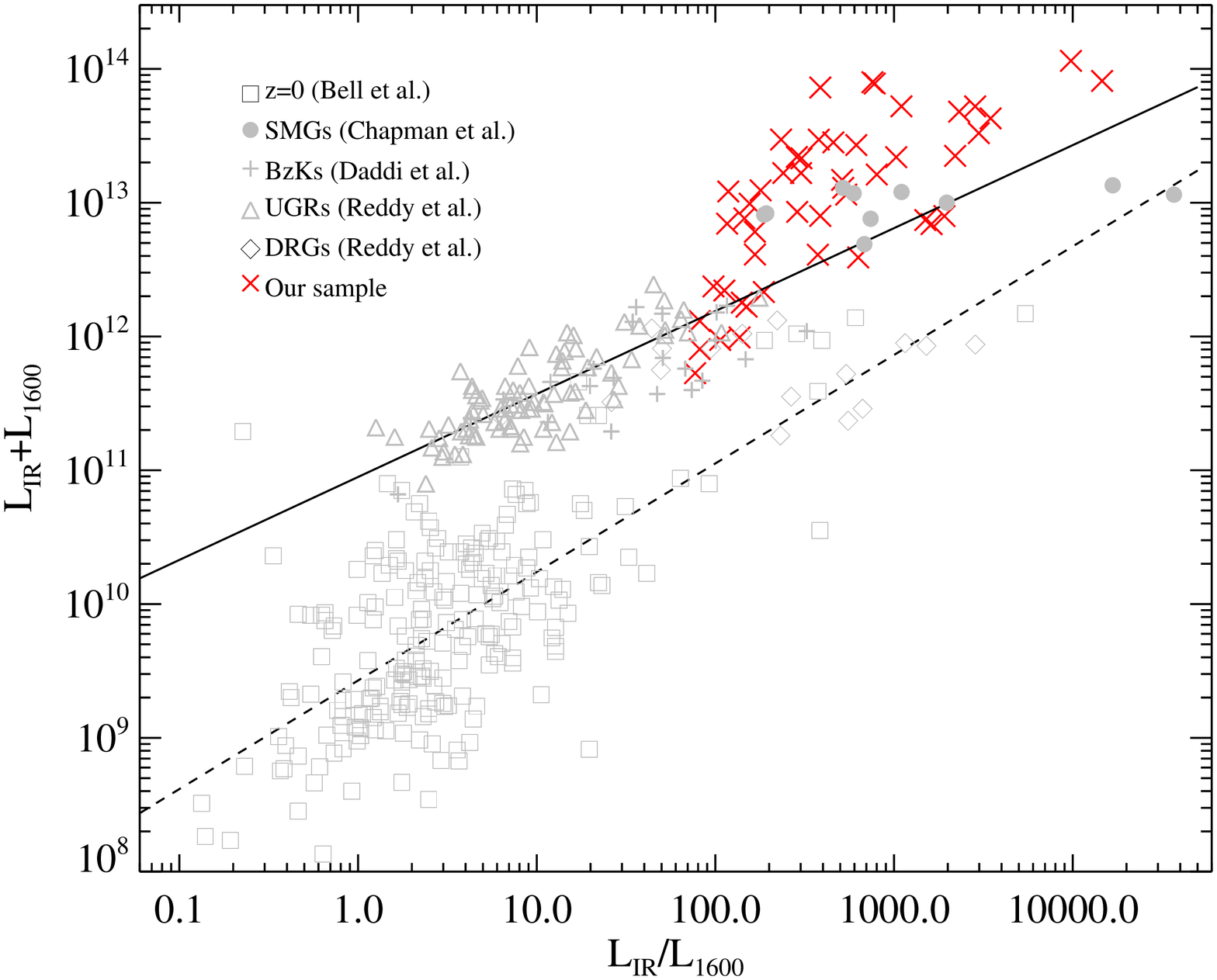}
\end{center}
\caption{Our sample compared with a number of other well-known samples including: UV-selected (UGRs), near-IR selected (BzKs, DRGs), far-IR-selected (SMGs), and near-IR-selected (DRGs). The other survey data compilation is based on Fig.~11 in Reddy et al. (2006).  To derive the values for our sample we assumed  a flat UV-slope, and $\log(L_{\rm{IR}})$\,=\,$\log L_{5.8}$\,+\,1.4. The solid line shows the best-fit relation for UGR galaxies (Reddy et al. 2006), while the dashed line shows the best-fit line for the $z$\,=\,0 sample. 
\label{uv-lir}}
\end{figure}

Lastly, we examine the rest-frame UV versus IR luminosities, $L_{\rm{1600}} / L_{\rm{IR}}$, which indicates how much UV optical photons being absorbed and reprocessed by dust, thus the averaged dust obscuration. We compare our sample with other well-known UV-selected or IR-selected high-$z$ populations. The derivation of $\log L_{\rm{IR}}$ for our sample was discussed in \S~\ref{contlums}. We derive the UV luminosities from the observed $R$-band photometry and redshifts, by assuming a flat  UV spectrum (i.e. $\nu F_{\nu}$\,=\,const, $\beta = -1$, for $\lambda$\,$<$\,4000\AA ).  This assumption of $\beta = -1$ is not too far off since SMG and LBG have the median $\beta$ value of -1.5 and -1 respectively \citep{chapman04, naveen06}. In addition, since at $z\sim2$, the {\sl R}-band samples the rest-frame 2333\AA, the K-correction in the estimate of $L_{\rm{1600}}$ from $R$-band photometry should be small. Most of our sources have $L_{\rm{1600}} \simlt 3\times10^{10}L_\odot$. Figure~\ref{uv-lir} shows $L_{\rm{IR}}/L_{\rm{1600}}$ vs. $L_{\rm{IR}}$\,+\,$L_{1600}$ for our sample, in comparison with other galaxy samples compiled by Reddy et al. (2006). The lower luminosity, lower redshift sub-sample of our galaxies overlap the most-luminous LBGs (a.k.a. UGRs) and the BzKs. The bulk of our sample however are closer to SCUBA galaxies, although our galaxies  are often more luminous. The mean $L_{IR}/L_{1600}$ ratio is about 5 for local normal galaxies, and much higher for high-z galaxies, with (1-100) for LBGs+BzKs and $100-3000$ for SMGs and our sample. Figure~\ref{uv-lir} suggests that more luminous galaxies tend to have higher IR/UV luminosity ratio, and on the whole our sources are among the most luminous and most obscured sources known. In particular, the local galaxy sample (open squares, Bell 2003) and high-z galaxies seem to follow different correlation between $L_{tot}$ vs. $L_{IR}/L_{1600}$ (dashed and solid lines). At a fixed IR-to-UV luminosity ratio, high-z galaxies are more luminous than $z=0$ counterparts, and at a fixed total luminosity, high-z galaxies have lower $L_{IR}/L_{1600}$ ratio than that of local ones (although local galaxies with very high IR-to-UV ratios are very few).

\section{Discussions and Summary}

\subsection{Nature of the PAH emitting ULIRGs at $z\sim2$}

The detection of PAH molecules in one third of the sample provides the direct evidence that aromatic features are already an important component of galaxy spectra at the early cosmological epoch. Quantitatively, the total energy in PAH features is roughly 20-70\%\ of the 5.8\um\ continuum luminosity ($\nu L_\nu (5.8\mu m)$), corresponding to $\sim$\,1\,--\,3\%\ of the total infrared luminosity. Studies utilizing IR SEDs need to properly consider the strength and distribution of PAH features since they have significant impact on the model predictions, such as redshift distribution and luminosity functions of 24\um\ selected samples \citep{lagache04}. In addition, the existence of PAH molecules in $z$\,$\sim$\,2 ULIRGs implies that metals from massive young stars formed at earlier epoch have gone through a complex process ---  cycling through dense molecular gas, forming photo-dessociation region (PDR) and starting another intense star formation again.  

The estimated rest-frame 5.8\um\ luminosities for our strong PAH sources are a factor of 5\,--\,10 times more luminous than the local starburst dominated ULIRGs. There are two possible physical explanations for this luminosity increase. The first scenario is that these $z\sim2$, strong PAH sources have much higher star formation efficiency as well as higher gas-to-mass fraction than local ULIRGs. The efficient star formation formation converts the large gas reservoir to stars and produce higher luminosity. This hypothesis is supported by studies of $z\sim2$ sub-mm galaxies, which have found that SMGs are scaled-up version of local ULIRGs, in the sense of luminosity, mass, star formation rate and gas fraction \citep{chapman04,greve05,tacconi06}. It is possible that the $z\sim2$ \spitz ULIRGs and SMGs are like local ULIRGs, and are in the stage of  ``maximum starburst'', {\it i.e.} an extreme mode of star formation where gas mass density is so high that it is balancing out the momentum-driven wind (radiative pressure, supernova explosion and stellar winds) \citep{Elmegreen99, Murray05}, resulting in the most efficient star formation. The second possible explanation is that although these $z\sim2$, PAH emitting ULIRGs are dominated by starburst at the mid-IR,  they could still have higher AGN contributions than the local starburst ULIRGs. Our spectra, with low spectral resolution and limited wavelength coverage, can not revealed the AGN observational signatures, such as high ionization ionic lines.  

\subsection{Nature of the non-PAH sources}

The AGN unification model as well as the observed hard X-ray background \citep{antonuci93} have
predicted the existence of type II quasars, with a rough ratio of type II to I of (4-10):1 \citep{madau94,comastri95}. However, UV and X-ray observations have so far failed to find many \citep{dan02}.  Our AGN dominated sources have $L_{5.8} \simgt 9\times10^{11}L_\odot = 3\times10^{45}$\,\ergs, corresponding to a total infrared luminosity on the order of $10^{47}$\,\ergs. Such a high luminosity is comparable to the bolometric luminosity of bright quasars. If most of this energy is generated by dust re-processing of the UV, soft X-ray photons from a black hole accretion disk, the inferred black hole mass is on the order of $10^{8}M_\odot$, assuming Eddington accretion rate. Even though one third of the sample do not show much mid-IR absorption along the line-of-sight, all of our sources are surrounded by substantial dusty material, as indicated by their large rest-frame IR to UV luminosity ratio. The deep silicate absorption systems probably have dust grain distribution with high filling factor, whereas the objects with weak absorption features have lumpy dust distributions.  Detailed modeling of deep silicate absorption spectra by Levenson et al. (2006) suggests that these sources contain central nuclear sources deeply embedded in a smooth distribution of dust that is geometrically and optically thick. 

Enormous infrared luminosity and high dust content suggest that our AGN sources are potentially type II quasars or at earlier stages before becoming supermassive black holes.  Of the total 52 sources, 22 are AGN dominated systems at $1.75 < z < 2.7$. This gives the co-moving volume density of $5\times10^{-7}Mpc^{-3}$, for the survey area of 3.7 sq. degree.  The type I quasar  co-moving space density at $z\sim2$ is about $6\times10^{-7}Mpc^{-3}$ with $\nu L_\nu (8\mu m) \simgt 4.1\times10^{45}$\,\ergs $\sim 10^{12}L_\odot$ \citep{brown06}.  This luminosity limit is roughly comparable to the averaged $L_{5.8}$ of our sample.  The space density measured from our sample is only a lower limit to all of type II quasars.  At face value, these space density estimates are consistent with the speculation that our AGN dominated systems could be in the early stages of quasars evolution. 

The AGN sources in our sample also have very red mid-IR (5-15\um) slope, with $\alpha \simgt 2$ ($f_\nu \propto \nu^{-\alpha}$), much redder than the averaged slope of $1.3$ for Palomar-Green (PG) quasars at $z\sim 0 - 1.2$ \citep{haas03}. The IRS spectra of our sample demonstrate that mid-IR colors (rest-frame, broad band) can not separate high-$z$ PAH emitting starbursts from AGNs with red mid-IR continua. For example, studies using blue 70\um-to-24\um\ color to select AGNs will miss a population of AGNs with red mid-IR continua, similar to our  sources. 

\subsection{Relation with other high-z galaxy populations}

We compare our sample directly with the Houck et al. (2005) \spitz selected $z\sim2$ sample. The IRS GTO team obtained low resolution IRS spectra for 58 sources with $f_\nu$(24\um)\,$\simgt$\,0.75\,mJy and $I > 24$mag (vega) in the Bootes field. Almost all of their observed spectra have deep silicate absorption/pure continuum only and {\it no} obvious, strong PAH emission, except a few spectra with broad peaks at $7-8$\um\ \citep{houck05, weedman06b}. Of their 58 sources, 76\%\ (44) have redshifts based on the IRS spectral features and the remaining 24\%\ are pure power-law without yielding any $z$ measurements. The results from the IRS GTO study \citep{houck05, weedman06b} differ from ours in the sense that they detect much less fraction of sources with strong PAH emission. This difference can be explained by the difference in the target selections. The IRS GTO study uses only two criteria: 1). $f_{24\mu m} > 0.75$mJy, 2). $\nu f_\nu(24\mu m)/\nu f_\nu(I) > 100$, and a few sources have $\nu f_\nu(24\mu m)/\nu f_\nu(I) > 75 - 100$.  

\begin{figure}[!ht]
\begin{center}
\plotfiddle{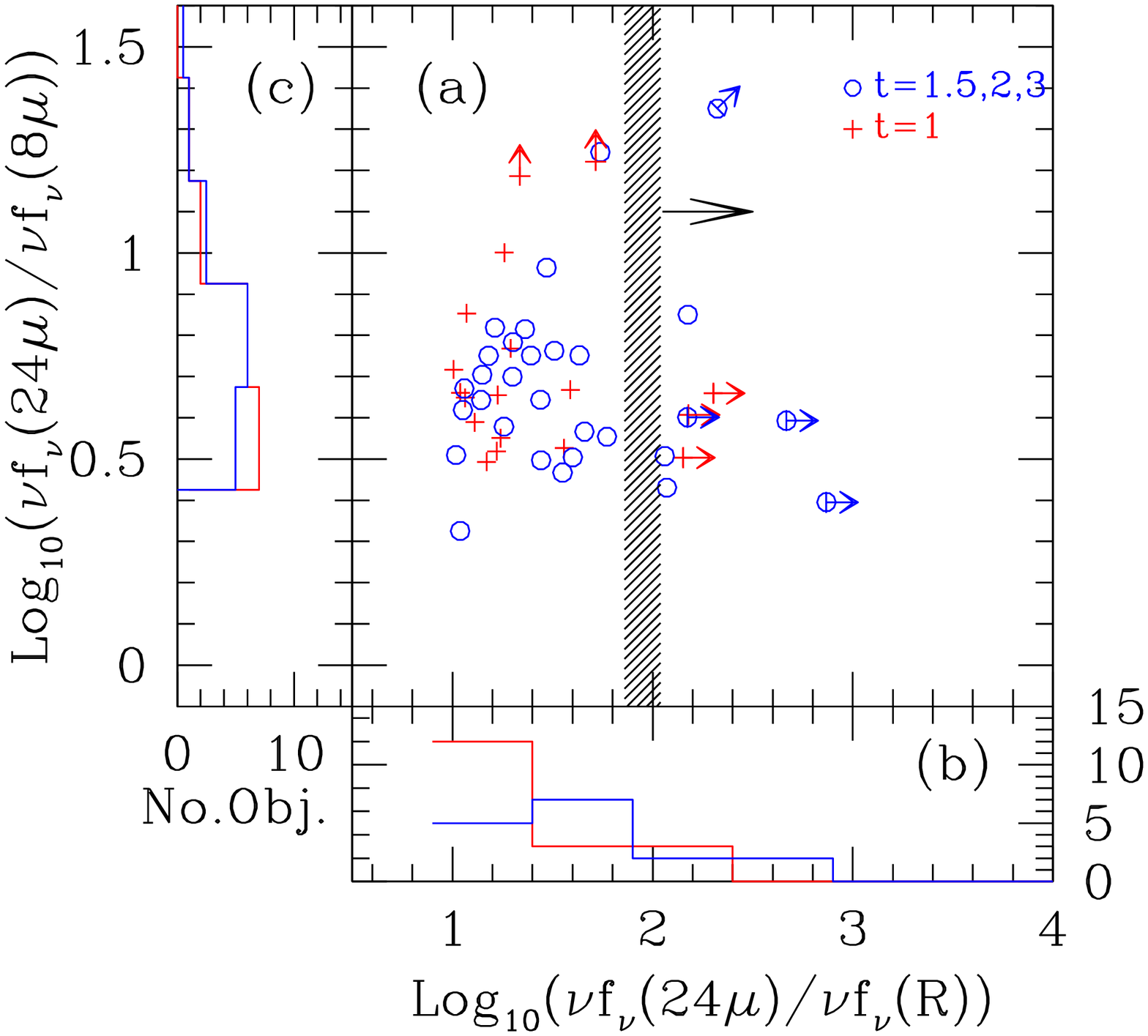}{3.5in}{0}{50}{50}{-170}{-90}
\end{center}
\caption{Panel (a) shows the 24/R and 24/8\um\ colors for the 52 sources we obtained the IRS low resolution spectra. The arrows indicate the sources with flux limits in R and/or 8\um\ band. The red plus symbols indicate the sources with strong PAH ($t=1$), and the blue open circles are for the deep absorption only spectra and mid-IR continuum dominated spectra ($t=1.5,2,3$). The shaded region marks the 24/R color cut used by the IRS GTO study \citep{houck05}. The majority of the GTO sources have 24/R color redder than the color marked by the shaded area. Panel (b) and (c) are the histograms of the source distributions for 24/R and 24/8\um\ respectively. The red and blue lines are for sources with and without strong PAH emission respectively.}
\label{colortype}
\end{figure}

Figure~\ref{colortype}, in three panels, to illustrate 24/8 and 24/R color-color diagram and the number of sources as functions of the two colors. Here the red and blue colors indicate objects with and without strong PAH respectively. The color is computed as $\nu f_\nu$ ratio between two wavelengths.  Both Panel (a) and (b) show that sources without strong PAH tend to have redder 24/R colors (blue lines) in comparison with sources with strong PAH (red). 

The Houck et al. sample has much redder (more than a factor of 10 in flux ratio) mid-IR to optical colors than our sample, occupying the regions redder than the shaded area in Panel (a), which indicates the GTO 24\um-R flux ratio limit.  Here we convert the GTO $I$ magnitude to $R$ used by our sample assuming $\nu f_\nu \propto$ constant (flat spectrum).  Most of our PAH strong sources lie the left side of the shaded area. The extremely high IR-to-optical ratio selects galaxies whose central energy engine is completely obscured by dust, very little UV photons escape from the centers, and furthermore, there is very little star formation outside the nuclear regions.  In contrast, some of our PAH emitting galaxies do have star formation outside of heavily obscured central regions.  

What about our sample in relation with other optical UV and (sub)-mm selected sample?
The $z\sim2$ ULIRGs revealed in our study represent an IR luminous population, which have very different dust and energetic properties from those of SMGs and LBGs. \spitz\ observations indicate that SMGs tend to have much fainter 24\um\ fluxes (on average a few 100$\mu$Jy) and their mid-IR spectra have strong PAH emission
\citep{lutz05a, pope06}. This is expected since \spitz 24\um\ selection tends to bias toward mid-IR bright, warmer,  AGN type sources whereas the sub-mm selection prefers galaxies containing substantial cold gas emitting at the far-IR \citep{blain04}. The initial 1.2mm obsesrvations of a subset of our sources by Lutz et al. (2005) also suggests that \spitz $z\sim2$ sources have on average less cold dust emission than SMGs. In addition, our $z\sim2$ sources have very little overlap with UV-optical selected Lyman break galaxies (LBG), which are much less infrared luminous than our sources \citep{naveen06}.  Of the sample of LBGs ($z\sim1.5-2.6$) in the HDFN, 60\%\ have 24\um\ counterparts with the average flux density of (20-30)$\mu$Jy, corresponding to $5\times10^{11}L_\odot$ \citep{naveen06}. Deep 24\um\ images reaching micron-Jy limits will probe the same galaxy population as SMGs and will have significant overlap with LBGs. 

We make crude estimates of space density of bright 24\um\ sources at $z\sim2$, in comparison with other types of galaxies. Of all the 24\um\ sources brighter than 0.9mJy over 3.7\,deg$^{2}$ in the FLS, 5\%\ (59) meet our sample selection criteria (our sample has 52 of these 59). Of the 52 sources from our sample, we have 34 at  \zr{1.5}{2.7}. Using this fraction to scale to the total 59 sources, the comoving space density of $z\sim2$ ULIRGs satisfied our selection criteria is  $n$\,$\sim$\,2\,$\times$\,$10^{-6} {\rm Mpc^{-3}}$ for $z = 2.1\pm0.6$ and $L_{\rm{IR}}$\,$\ge$\,5\,$\times$\lir{12} (see Figure~\ref{lumtype} for the limit). This volume density is only for ULIRGs with very red mid-IR color , and obviously is only a lower limit for the total population of $z\sim2$ infrared luminous galaxies.  For comparison, the comoving number density of sub-mm detected galaxies at $\langle$$z$$\rangle$\,=\,2.2 is roughly $6\times 10^{-6}{\rm Mpc^{-3}}$ for $L_{\rm{IR}}$\,$\ge$\,4\,$\times$\,\lir{12} \citep{scott04}. This implies that our survey has revealed a population of $z\sim2$ ULIRGs which is as numerous as SMGs, and the total infrared luminous population at $z\sim2$, including both ours (warm) and SMGs (cold), is at least $10^{-5}{\rm Mpc^{-3}}$ for $L_{\rm{IR}}$\,$\ge$\,4-5\,$\times$\,\lir{12}. 

The rest-frame UV color selected galaxies have co-moving space density of 2\,$\times$\,$10^{-3}{\rm Mpc^{-3}}$ for the Balmer break selected galaxies at $\langle$$z$$\rangle$\,=\,1.77 and $\langle$$z$$\rangle$\,=\,2.32 \citep{adelberger04a}. Optical/UV selected galaxies are more numerous than IR luminous galaxies (\spitz + SMGs) by a factor of 200, and they populate the steep, faint end of the infrared luminosity function, with $L_{IR} \simlt 10^{11}L_\odot$. However,  IR luminous galaxies, including SMGs and ones from our sample, are probably more luminous by a factor of $>100$.  Therefore, the integrated luminosity density from SMGs and \spitz selected ULIRGs at $z\sim2$ 
is roughly comparable to that from UV selected galaxies. 

\section{Summary}

We have used the IRS on \spitz\ to observe a sample of 52, 24\um\ brighter than 1mJy galaxies in the XFLS. Of the 52 targets in our sample, 47 (90\%) have measurable redshifts based on the IRS spectra, with 35 at $1.5<z<3.2$, and 12 at $0.62< z<1.3$. The observed spectra fall crudely into three categories, one third with strong PAH emission, one third with only deep silicate absorption and the remaining third of the sample with strong mid-IR continuum plus weak emission and/or absorption features. The sources in our sample are very luminous, with estimated 5.8\um\ ($\nu L_\nu(5.8\mu m)$) continuum luminosity $\simgt 10^{12} L_\odot$, implying $L_{\rm IR}$ on the order of $\simgt 10^{13}L_\odot$. Our sources, by selection, have very high infrared to the rest-frame UV luminosity ratios ($L_{\rm {IR}}/L_{1600\AA} \sim 100-1000$), and are, thus heavily dust obscured in the UV. At the mid-IR 10\um, $\sim$60\%\ of our sample have optical depth $\tau_{9.8\mu m}$ greater than 1, implying high dust obscuration along line-of-sight. Over all, our sources are among the most luminous, and on average, the most dust obscured at $z\sim2$.  Our survey revealed an infrared luminous population at $z\sim2$, which has very little overlap with SMGs and other optical/near-IR selected galaxies. 

For the third of the sample with PAH emission, we estimate the total energy in the PAH features (continuum subtracted) is roughly 20-70\%\  of the 5.8\um\ continuum luminosity ($\nu L_\nu (5.8\mu m)$), corresponding to $\sim$\,1\,--\,3\%\ of the total infrared luminosity. This implies that PAH features are important components of mid-IR SEDs at high redshift. On average, our PAH emitting ULIRGs are about a factor of (5-10) more luminous than starburst ULIRGs at $z\sim0$. The higher luminosity from these systems could be due to the combination of very efficient star formation and higher fraction of gas-to-mass ratio at $z\sim2$.  Our study has revealed a population of extremely IR luminous galaxies at $z\sim2$, whose physical properties are diverse, with a third of the sample powered by dusty star formation, one third of sample being deeply embedded systems, whose dust heating sources are heavily obscured by dust and their energetic nature can not be determined unambiguously by the present dataset alone. Finally, the remainder third of the sample are probably AGN powered systems. The enormous energy output and high fraction of dust content make these sources prime candidates for long-sought type II quasars. 

\acknowledgements
We would like to thank many helpful discussions with Henrik Spoon, Tom Soifer, 
and Guilaine Lagache.  Support for this work was provided by NASA through 
an award issued by JPL/Caltech. A. Sajina acknowledges support through
NASA grant 09865 from the Space Telescope Science Institute, which is 
operated by the Association of Universities for Research in Astronomy, Inc., 
under NASA contract BAS5-26555. 

\clearpage

\clearpage


\clearpage
\begin{deluxetable}{rcccccccc}
\tabletypesize{\scriptsize}
\tablecaption{IRS Low Resolution Spectroscopy Observation Log \label{target_tab}}
\tablewidth{0pt}
\tablehead{
\colhead{MIPS ID} & 
\colhead{IAU ID} &
\colhead{$f(24\mu m)$} &
\colhead{$f(8\mu m)$} &
\colhead{$R$\tablenotemark{a}} &
\colhead{SL\tablenotemark{b} 1st} &
\colhead{LL\tablenotemark{b} 2nd} &
\colhead{LL 1st}  \\
   & & mJy & $\mu$Jy & $\mu$Jy & seconds & seconds & seconds
}
\startdata
MIPS22699 & SST24 J172047.47+590815.1 & 906.769 & 94.840 & 0.186 & \nodata & 7x2x122 & 9x2x122 \\
MIPS8493 &  SST24 J171805.11+600832.8 & 916.846 & 30.514 & 1.474 & \nodata & 8x2x122 & 8x2x122 \\
MIPS506 &   SST24 J171138.59+583836.7 & 920.141 & 20.000 & 1.237 & \nodata & 8x2x122 & 8x2x122 \\
MIPS22661 & SST24 J171819.64+590242.6 & 951.655 & 79.439 & 0.186 & \nodata & 6x2x122 & 8x2x122 \\
MIPS22651 & SST24 J171926.47+590929.9 & 960.877 & 79.083 & 0.186 & \nodata & 6x2x122 & 8x2x122\\
MIPS429 &   SST24 J171611.81+591213.3 & 999.243 & 20.000 & 0.560 & \nodata & 6x2x122 & 8x2x122 \\
MIPS464 &   SST24 J171439.57+585632.1 & 955.156  & 56.454 & 1.128 & \nodata & 6x2x122 & 8x2x122 \\
MIPS16144 & SST24 J172422.10+593150.8 & 1010.148 & 94.712 & 1.692 & \nodata & 6x2x122 & 8x2x122 \\
MIPS16122 & SST24 J172051.48+600149.1 & 1034.885 & 93.621 & 0.661 & \nodata & 6x2x122 & 8x2x122 \\
MIPS22600 & SST24 J172218.34+584144.6 & 1039.513 & 76.647 & 1.805 & \nodata & 6x2x122 & 8x2x122 \\
MIPS16113 & SST24 J172126.42+601646.1 & 1042.892 & 37.765 & 1.029 & \nodata & 6x2x122 & 8x2x122 \\
MIPS22582 & SST24 J172124.58+592029.5 & 1058.413 & 27.807 & 0.186 & \nodata & 6x2x122 & 8x2x122 \\
MIPS16095 & SST24 J172359.74+595752.1 & 1066.319 & 85.501 & 2.757 & \nodata & 6x2x122 & 8x2x122 \\
MIPS16080 & SST24 J171844.77+600115.9 & 1097.299 & 64.978 & 2.111 & \nodata & 6x2x122 & 8x2x122 \\
MIPS22558 & SST24 J172045.17+585221.4 & 1107.165 & 63.761 & 1.001 & \nodata & 6x2x122 & 8x2x122 \\
MIPS22554 & SST24 J172059.80+591125.7 & 1111.963 & 51.993 & 2.757 & \nodata & 6x2x122 & 8x2x122 \\
MIPS8342 &  SST24 J171411.55+601109.3 & 1112.190 & 117.992 & 1.171 & \nodata & 6x2x122 & 8x2x122 \\
MIPS16059 & SST24 J172428.44+601533.2 & 1137.134 & 86.168 & 1.204 & \nodata & 6x2x122 & 8x2x122 \\
MIPS8327 &  SST24 J171535.78+602825.5 & 1144.055 & 58.435 & 1.447 & \nodata & 6x2x122 & 8x2x122 \\
MIPS22530 & SST24 J172303.30+591600.2 & 1156.188 & 65.766 & 1.724 & \nodata & 6x2x122 & 8x2x122 \\
MIPS16030 & SST24 J172000.32+601520.9 & 1192.988 & 120.459 & 2.092 & \nodata & 6x2x122 & 8x2x122 \\
MIPS16006 & SST24 J171901.58+595418.7 & 1235.011 & 102.847 & 0.833 &\nodata & 8x2x122 & 8x2x122 \\
MIPS8268 &  SST24 J171624.74+593752.6 & 1253.351 & 197.470 & 3.346 & \nodata & 9x2x122 & 5x2x122 \\
MIPS22482 & SST24 J172100.39+585931.0 & 1272.742 & 91.189 & 0.956 & \nodata & 6x2x122 & 8x2x122 \\
MIPS289 &   SST24 J171350.00+585656.8 & 1280.379 & 93.548 & 0.186 & \nodata & 6x2x122 & 8x2x122 \\
MIPS15977 & SST24 J171855.67+594545.4 & 1297.162 & 114.200 & 2.092 & \nodata & 6x2x122 & 8x2x122 \\
MIPS283 &   SST24 J171458.27+592411.2 & 1301.197 & 139.303 & 2.561 & \nodata & 6x2x122 & 8x2x122 \\
MIPS279 &   SST24 J171217.04+584810.4 & 1305.079 & 121.469 & 0.643 & \nodata & 8x2x122 & 8x2x122 \\
MIPS22467 & SST24 J171845.47+583913.4 & 1310.665 & 135.130 & 3.668 & \nodata & 8x2x122 & 8x2x122 \\
MIPS15958 & SST24 J172324.21+592455.7 & 1341.651 & 165.844 & 0.332 & 1x2x242 & 5x2x122 & 7x2x122 \\
MIPS8245 &  SST24 J171536.34+593614.8 & 1344.083 & 20.000 & 0.186 & \nodata & 6x2x122 & 8x2x122 \\
MIPS8242 &  SST24 J171433.17+593911.2 & 1356.030 & 68.640 & 2.424 & \nodata & 8x2x122 & 8x2x122 \\
MIPS15949 & SST24 J172109.22+601501.3 & 1386.653 & 76.137 & 2.016 & \nodata & 8x2x122 & 8x2x122 \\
MIPS15929 & SST24 J172406.95+592416.3 & 1446.681 & 61.511 & 0.186 & \nodata & 6x2x122 & 8x2x122 \\
MIPS15928 & SST24 J171917.45+601519.9 & 1455.662 & 108.901 & 3.668 & \nodata & 8x2x122 & 8x2x122 \\
MIPS8207 &  SST24 J171448.54+594641.1 & 1466.097 & 93.747 & 4.212 & \nodata & 8x2x122 & 8x2x122 \\
MIPS227 &   SST24 J171456.24+583816.2 & 1482.509 & 105.391 & 3.771 & \nodata & 8x2x122 & 8x2x122 \\
MIPS22404 & SST24 J172151.77+585327.7 & 1497.993 & 108.952 & 3.986 & \nodata & 6x2x122 & 8x2x122 \\
MIPS8196 &  SST24 J171510.28+600955.2 & 1500.852 & 98.800 & 3.108 & \nodata & 8x2x122 & 8x2x122 \\
MIPS8184 &  SST24 J171226.76+595953.5 & 1540.086 & 132.193 & 3.471 & \nodata & 6x2x122 & 8x2x122 \\
MIPS15880 & SST24 J172119.46+595817.2 & 1609.199 & 95.078 & 1.088 & \nodata & 6x2x122 & 7x2x122 \\
MIPS180 &   SST24 J171543.54+583531.2 & 1683.941 & 31.977 & 0.896 & \nodata & 8x2x122 & 8x2x122 \\
MIPS8135 &  SST24 J171455.71+600822.6 & 1769.055 & 103.991 & 0.294 & \nodata & 8x2x122 & 8x2x122 \\
MIPS15840 & SST24 J171922.40+600500.4 & 1820.508 & 188.874 & 0.462 & 1x2x242 & 6x2x122 & 7x2x122 \\
MIPS133 &   SST24 J171433.68+592119.3 & 1989.473 & 150.865 & 4.173 & 1x2x242 & 2x2x122 & 2x2x122 \\
MIPS22303 & SST24 J171848.80+585115.1 & 2029.962 & 95.502 & 0.395 & \nodata & 2x2x122 & 3x2x122 \\
MIPS22277 & SST24 J171826.67+584242.1 & 2298.740 & 227.975 & 1.856 & 1x2x242 & 2x2x122 & 3x2x122 \\
MIPS110 &   SST24 J171215.44+585227.9 & 2300.304 & 261.559 & 1.890 & 1x2x242 & 2x2x122 & 3x2x122 \\
MIPS8034 &  SST24 J171210.28+601858.1 & 2909.007 & 193.751 & 4.251 & 1x2x242 & 2x2x122 & 3x2x122 \\
MIPS78 &    SST24 J171538.18+592540.1 & 2972.799 & 252.988 & 0.186 & 1x2x242 & 2x2x122 & 3x2x122 \\
MIPS22204 & SST24 J171844.38+592000.5 & 4100.928 & 428.290 & 2.996 & 1x2x242 & 2x2x122 & 3x2x122 \\
MIPS42 &    SST24 J171758.44+592816.8 & 4711.865 & 632.269 & 0.186 & 1x2x242 & 2x2x122 & 3x2x122 \\
\enddata
\tablenotetext{a}{For 8\um\ none-detected sources, we list 2$\sigma$ flux limits, taken
from the XFLS IRAC catalog by \citet{lacy05}.}
\tablenotetext{b}{SL and LL stand for Short-Low and Long-Low module of the IRS. The total exposure time is equal to No.of repeats
$\times$ 2 (No. of dither position)
$\times$ individual exposure time.}
\end{deluxetable}

\begin{deluxetable}{rllcll}
\tablecaption{The Measured Redshifts and Types for Our Sample \label{ztable}}
\tablewidth{0pt}              
\tablehead{              
\colhead{MID} &             
\colhead{$z_{vis}\pm 1\sigma$} &            
\colhead{$\log L_{5.8}$} &            
\colhead{$q_z$} &             
\colhead{Type} &             
\colhead{features} \\             
}              
\startdata         
MIPS22404 & $0.61\pm0.06$ & 10.32$\pm$0.07 & a & 1 & 9.8,11.3,12.8 \\
MIPS22554 & $0.82\pm0.05$ & 10.50$\pm$0.09 & a & 1 & 8.6,9.8,11.3,12.8 \\
MIPS8207 & $0.84\pm0.03$ & 10.71$\pm$0.07 & a & 1 & 9.8,11.3,12.8 \\
MIPS22600 & $0.86\pm0.03$ & 10.59$\pm$0.10 & a & 1 & 9.8,11.3,12.7,15.5 \\
MIPS283 & $0.94\pm0.04$ & 10.85$\pm$0.08 & a & 1 & 9.8,11.3,12.8 \\
MIPS16030 & $0.98\pm0.04$ & 10.93$\pm$0.07 & a & 1 & 9.8,11.3,12.8 \\
MIPS8184 & $0.99\pm0.02$ & 10.94$\pm$0.06 & a & 1 & 9.8,11.3,12.8,15.5 \\
MIPS15928 & $1.52\pm0.02$ & 11.44$\pm$0.05 & a & 1 & 6.2,7.7,8.6,11.3,12.8 \\
MIPS22651 & $1.73\pm0.02$ & 11.50$\pm$0.15 & a & 1 & 6.2,7.7,8.6,11.3,12.8 \\
MIPS22277 & $1.77\pm0.03$ & 12.03$\pm$0.08 & a & 1 & 6.2,7.7,9.8 \\
MIPS8493 & $1.80\pm0.01$ & 11.38$\pm$0.23 & a & 1 & 7.7,9.8,11.3,12.7 \\
MIPS22482 & $1.84\pm0.02$ & 11.71$\pm$0.08 & a & 1 & 7.7,8.6,9.8,11.3 \\
MIPS289 & $1.86\pm0.01$ & 11.47$\pm$0.23 & a & 1 & 6.2,7.7,8.6,9.8,11.3 \\
MIPS22530 & $1.96\pm0.01$ & 11.48$\pm$0.25 & a & 1 & 6.2,7.7,8.6,9.8,11.3 \\
MIPS429 & $2.09\pm0.02$ & 11.66$\pm$0.09 & a & 1 & ArII,7.7,8.6,9.8 \\
MIPS16144 & $2.13\pm0.02$ & 11.59$\pm$0.28 & a & 1 & 6.2,7.7,8.6,9.8,11.3 \\
MIPS506 & $2.52\pm0.02$ & 11.82$\pm$0.20 & a & 1 & 6.2,7.7,9.8 \\
\hline
MIPS15880 & $1.64\pm0.05$ & 11.94$\pm$0.03 & a & 1.5 & 7.7,9.8 \\
MIPS464 & $1.85\pm0.07$ & 11.53$\pm$0.06 & a & 1.5 & 7.7 \\
MIPS16113 & $1.93\pm0.07$ & 11.77$\pm$0.08 & a & 1.5 & 7.7,9.8 \\
MIPS42 & $1.95\pm0.07$ & 12.51$\pm$0.02 & a & 1.5 & 7.7,9.8 \\
MIPS16122 & $1.97\pm0.05$ & 11.81$\pm$0.06 & a & 1.5 & 7.7,9.8,11.3 \\
MIPS16080 & $2.04\pm0.06$ & 11.93$\pm$0.04 & a & 1.5 & 7.7,9.8 \\
MIPS22204 & $2.08\pm0.03$ & 12.50$\pm$0.02 & a & 1.5 & 7.7,9.8,11.3 \\
MIPS16059 & $2.43\pm0.07$ & 11.69$\pm$0.05 & a & 1.5 & 6.2,7.7,9.8 \\
MIPS180 & $2.47\pm0.04$ & 12.32$\pm$0.03 & a & 1.5 & 6.2,7.7,9.8 \\
MIPS8327 & $2.48\pm0.06$ & 12.07$\pm$0.07 & a & 1.5 & 6.2,7.7,9.8 \\
MIPS8196 & $2.6\pm0.1$ & 12.46$\pm$0.02 & a & 1.5 & 7.7,9.8 \\
MIPS8242 & $2.45\pm0.04$ & 12.07$\pm$0.04 & a & 1.5 & 6.2,7.7,8.6,9.8 \\
MIPS22699 & $2.59\pm0.04$ & 12.12$\pm$0.04 & a & 2 & 6.2,7.7,8.6,9.8 \\
MIPS78 & $2.65\pm0.1$ & 12.66$\pm$0.02 & a & 2 & 9.8 \\
MIPS22303 & $2.34\pm0.14$ & 12.32$\pm$0.06 & a & 2 & 7.7,9.8 \\
MIPS8245 & $2.7\pm0.1$ & 12.23$\pm$0.08 & a & 2 & 9.8 \\
MIPS22558 & $3.2\pm0.1$ & 12.49$\pm$0.02 & a & 2 & 7.7,9.8 \\
\hline
MIPS8342 & $1.57\pm0.09$ & 11.50$\pm$0.08 & a & 3 & 6.2,7.7,11.3,12.8 \\
MIPS22661 & $1.75\pm0.03$ & 11.44$\pm$0.07 & a & 3 & 7.7,11.3 \\
MIPS16095 & $1.81\pm0.05$ & 11.69$\pm$0.05 & a & 3 & 9.8,11.3,12.8 \\
MIPS227 & $1.87\pm0.05$ & 11.68$\pm$0.06 & a & 3 & 7.7,9.8,11.3,12.8 \\
MIPS15958 & $1.97\pm0.05$ & 11.95$\pm$0.05 & a & 3 & 7.7,9.8 \\
MIPS15949 & $2.15\pm0.03$ & 11.94$\pm$0.04 & a & 3 & 7.7,11.3 \\
MIPS15840 & $2.3\pm0.1$ & 12.28$\pm$0.03 & a & 3 & weak 9.8\\
MIPS15977 & $1.85\pm0.07$ & 11.82$\pm$0.05 & b & 3 & weak 6.2,7.7,11.3 \\
MIPS110 & $1.0\pm0.1$ & 11.21$\pm$0.07 & b & 3 & 10\um\ \\
MIPS133 & $1.0\pm0.2$ & 10.97$\pm$0.05 & b & 3 & 10\um\ \\
MIPS22467 & $0.7\pm0.2$ & 10.57$\pm$0.05 & b & 3 & 9.8 \\
MIPS8268 & $0.8\pm0.2$ & 10.82$\pm$0.05 & b & 3 & 10\um\ \\
MIPS8034 & $0.7\pm0.2$ & 11.21$\pm$0.08 & b & 3 & 10\um\ \\

\hline 
MIPS279     & 999\tablenotemark{a} & 11857664  & a & 4 & \\             
MIPS22582 & 999 & 11870208 & a & 4 & \\     
MIPS15929 & 999 & 11865088 & a & 4 & \\     
MIPS16006 & 999 &  11856896 & a & 4 & \\     
MIPS8135  & 999 &  11861248 & a & 4 & \\     
\enddata 
\tablenotetext{a}{The value 999 means that no redshifts were derived based on the IRS spectra.}             
\end{deluxetable}              

\begin{deluxetable}{rcccc}
\tabletypesize{\scriptsize}
\tablecaption{The Keck Spectroscopy of the IRS Sample \label{keckspec}}
\tablewidth{0pt}
\tablehead{
\colhead{MID} &
\colhead{$z_{keck}$} &
\colhead{$z_{IRS}$} &
\colhead{Type(IRS)} &
\colhead{Type(Keck)} \\
}
\startdata
& & NIRSPEC Data & & \\
\hline
MIPS22530 & 1.9511 & 1.96 & strong PAH ($t=1$) & H$\alpha$ \\
MIPS8327 & 2.4426 & 2.48 & highly obscured ($t=1.5$) & H$\alpha$ \\
MIPS506 & 2.4691 & 2.52 & strong PAH ($t=1$) & H$\alpha$ \\
MIPS16059 & 2.3252 & 2.43 & highly obscured ($t=1.5$) & H$\alpha$ \\
MIPS22204 & 1.9707 & 2.08 & highly obscured ($t=1.5$) & H$\alpha$ \\
MIPS16080 & 2.0077 & 2.04 & highly obscured ($t=1.5$) & Sy2, [OIII]5007, H$\alpha$ \\
\hline
& & LRIS Data & & \\
\hline
MIPS8207 & 0.8343 & 0.84 & strong PAH ($t=1$) & Sy2, [OII]3727,[NeIII]3868 \\
MIPS133 & 0.9060 & 1.0 & weak features ($t=3$) & Sy2, [OII]3727,[NeIII]3868,[OIII]5007 \\
MIPS283 & 0.9374 & 0.94 & strong PAH ($t=1$) & Sy2, [OII]3727,[NeIII]3868 \\
MIPS227 & 1.6370 & 1.87 & weak features ($t=3$) & AGN, Ly$\alpha$,CIV],HeII,CIII],MgII \\
MIPS16030 & 0.9865 & 0.98 & strong PAH ($t=1$) & star forming, [OII]3727\\
MIPS16080 & 2.011 & 1.95 & highly obscured ($t=1.5$) & QSO, broad CII]2326 \\
\hline
& & DEIMOS Data & & \\
\hline
MIPS22404 & 0.6101 & 0.61 & strong PAH ($t=1$) & star forming, [OII]3727 \\
MIPS22600 & 0.8504 & 0.86 & strong PAH ($t=1$) & star forming, [OII]3727\\
MIPS22554 & 0.8242 & 0.82 & strong PAH ($t=1$) & star forming, [OII]3727 \\
MIPS279 & 1.2319 & 1.3 & weak features ($t=3$) & AGN \\
MIPS110 & 1.0505 & 1.0 & weak features ($t=3$) & AGN \\
\enddata
\end{deluxetable}

\end{document}